\RequirePackage[l2tabu, orthodox]{nag}
\RequirePackage{snapshot}

\documentclass[10pt,onecolumn]{extarticle}

\makeatletter
\if@twocolumn
  \usepackage[dvips,letterpaper,top=0.5in, bottom=0.5in, left=0.75in, right=0.5in,includefoot,heightrounded]{geometry}
\else
  \usepackage[dvips,letterpaper,margin=1in,includefoot,heightrounded]{geometry}
\fi

\usepackage{srcltx}

\iflanguage{portuges}
    {\newcommand{\keywordname}{Palavras-chaves}}
    {\newcommand{\keywordname}{Keywords}}

\usepackage{amsmath}
\usepackage{amssymb}
\usepackage{amsfonts}
\usepackage{mathtools}

\usepackage{abstract}

\usepackage{setspace}

\usepackage{multicol}

\usepackage{tabularx}
\usepackage{arydshln} %
\usepackage{booktabs} %
\usepackage{placeins}

\usepackage{hyperref}
\usepackage[normalem]{ulem}

\usepackage{graphicx}
\usepackage{multirow}
\usepackage{float}
\usepackage{psfrag}
\usepackage{subfig}

\usepackage{algorithm}
\usepackage{algorithmic}

\usepackage[short,12hr]{datetime}

       \usepackage{fouriernc}
\makeatletter

\newenvironment{rsmallmatrix}{\null\,\vcenter\bgroup
  \Let@\restore@math@cr\default@tag
  \baselineskip6\ex@ \lineskip1.5\ex@ \lineskiplimit\lineskip
  \ialign\bgroup\hfil$\m@th\scriptstyle##$&&\thickspace\hfil
  $\m@th\scriptstyle##$\crcr
}{%
  \crcr\egroup\egroup\,%
}

\makeatletter
\newcommand{\algrule}[1][.2pt]{\par\vskip.5\baselineskip\hrule height #1\par\vskip.5\baselineskip}
\makeatother

\newcommand{\printtitle}{%
\makeatletter
\if@twocolumn

\twocolumn[%
  \maketitle
  \begin{onecolabstract}
    \myabstract
  \end{onecolabstract}
  \begin{center}
    \small
    \textbf{\keywordname}
    \\\medskip
    \mykeywords
  \end{center}
  \bigskip
]
\saythanks
\else
  \maketitle
  \begin{onecolabstract}
    \myabstract
  \end{onecolabstract}
  \begin{center}
    \small
    \textbf{\keywordname}
    \\\medskip
    \mykeywords
  \end{center}
  \bigskip
  \onehalfspacing
\fi
\makeatother
}

\author{%
A.~P.~Rad\"unz%
\thanks{Independent researcher,
previously with the
Programa de Pós-Graduação em Estatística, Universidade Federal de Pernambuco, Recife, Brazil.}
\and
L.~Portella%
\thanks{
Universidade Federal de Santa Maria,
Cachoeira do Sul, Brazil;
previously with the
Programa de Pós-Graduação em Estatística, Universidade Federal de Pernambuco, Recife, Brazil.
}
\and
R.~S.~Oliveira%
\thanks{Independent researcher,
previously with the
Universidade Federal de Pernambuco (UFPE), Recife, Brazil.}
\and
F.~M.~Bayer%
\thanks{Departamento de Estat\'istica and LACESM, Universidade Federal de Santa Maria, Santa Maria, Brazil.}
\and
R.~J.~Cintra%
\thanks{%
Industrial Signal Processing Laboratory,
Departamento de Tecnologia,
UFPE,
Caruaru,
Brazil.
E-mail: \url{rjdsc@de.ufpe.br}}
}

\title{%
Extensions on Low-complexity DCT Approximations for Larger Blocklengths Based on Minimal Angle Similarity}

\newcommand{\myabstract}{%
The discrete cosine transform (DCT) is a central tool for image and video coding because it can be related to the Karhunen-Lo\`eve transform (KLT), which is the optimal transform in terms of retained transform coefficients and data decorrelation.
In this paper, we introduce 16-, 32-, and 64-point low-complexity
DCT
approximations by minimizing individually the angle between the rows of the exact DCT matrix and the matrix induced by the approximate transforms.
According to some classical figures of merit, the proposed transforms outperformed the approximations for the DCT already known in the literature. Fast algorithms were also developed for the low-complexity transforms, asserting a good balance between the performance and its computational cost.
Practical applications in image encoding showed the relevance of the transforms in this context.
In fact, the experiments showed that the proposed transforms had better results than the known approximations in the literature for the cases of $16$, $32$, and $64$ blocklength.
}

\newcommand{\mykeywords}{%
Discrete cosine transform, fast algorithms, image compression,
low-complexity transform
}

\date{}

\begin{document}

\printtitle


%

\section{Introduction}
\label{S:intro}

Current technological trends
suggest an ever-increasing
demand
for efficient, low-power, low-complexity
digital signal processing
methods~\cite{8416771,8575164,IBHAZE2020100055,9093864}.
In this context, many important discrete transforms have become useful tools for signal coding and data decorrelation~\cite{gonzalez2012,britanak2007discrete,ochoa2019discrete, poularikas2010transforms,salomon2007} such as the discrete Fourier transform (DFT), the discrete Hartley transform (DHT), the Walsh-Hadamard transform (WHT), the discrete Tchebichef transform (DTT), the discrete cosine transform (DCT), among others.
Data compression techniques~\cite{jain1981image,welch1984technique,penn1992,jolliffe2002pca}
address the problem of removing
redundancy from data~\cite[pg.~2]{salomon2007}.
Such removal
can be accomplished by using
the Karhunen-Lo\`eve transform (KLT)~\cite{ochoa2019discrete,britanak2007discrete}, which is the optimal
tool
in terms of energy compaction.
Indeed, the KLT packs the energy of the input signal in few
transform-domain coefficients and diagona\-li\-zes the covariance or correlation matrix of the data,
resulting in a completely decorrelated signal~\cite[pg.~10]{ochoa2019discrete}.
In practice,
the KLT is not widely adopted because
it is a data-dependent transformation,
which severely precludes the development of fast algorithms.
Fast algorithms can highly reduce the arithmetic cost of the transform since it searches for a computationally efficient way of implementing it~\cite[pg.~2]{blahut2010}.

However,
if the input signal is a first-order Markovian process,
then the KLT
matrix
depends only
on the correlation coefficient $\rho$ of the process.
Natural images
constitute
a representative class of such kind of data,
which often present high values of $\rho$.
When $\rho \to 1$,
the KLT becomes the type-II
discrete cosine transform
which is independent of the input data or any
other parameter~\cite[pg.~56]{britanak2007discrete}.
Although natural images do not necessarily present $\rho=1$,
their correlation coefficient is sufficiently high~\cite{gonzalez2012}
to make the DCT
extremely efficient and popular in image and video coding,
such as
JPEG~\cite{wallace1992jpeg},
MPEG~\cite{Puri2004},
and
HEVC~\cite{pourazad2012}.
Since its introduction in~\cite{ahmed1974}, the DCT has been shown to be a superior tool when compared
to the other transforms known in the literature in this context of image and video coding.
This fact was confirmed in a number of independent works such as~\cite{wallace1992jpeg,Clarke1981,clarke1983application,jain1981image,ochoa2019discrete,britanak2007discrete,rao1990,gonzalez2012,sayood2017introduction,rao2001transform}.
The fact that the DCT is data-independent allows the development of
low-complexity fast algorithms
for its calculation~\cite{blahut2010}.

Nevertheless,
under scenarios of
severe restrictions on processing power or energy autonomy~\cite{cintra2014low,sheltami2016data},
the arithmetic cost of computing the DCT
by traditional algorithms
might still be a hindrance.
Thus,
several
multiplierless
low-complexity DCT approximations
have been
developed
as detailed in~\cite{Haweel2001,Liang2001,cintra2011dct,cintra2014low,bas2008n,bas2010,tablada2015,Cintra2012,coutinho2015,oliveira2019low}.
Particularly, we separate the method presented in~\cite{oliveira2019low}, where a DCT approximation is derived by minimizing
the angle between each
corresponding rows
of the exact and approximate transforms.
The resulting 8-point DCT approximation introduced in~\cite{oliveira2019low} outperforms well-known DCT approximations in the literature according to
classical figures of merit
as the mean squared error, total energy error, coding gain, and transform efficiency. Furthermore,
the associated fast algorithm requires only 24~additions and
6~bit-shifting operations.

The search for 8-point DCT approximations is a relatively mature area of research since the 8-point DCT is a key block in many image and video processing applications.
However, with the advancement of image and video encoding technology,
there is a demand for larger blocklengths that could be employed in
modern codecs~\cite{shi1999image}.
For instance, there is a new video standard coding, the Versatile Video Coding (VVC)~\cite{zhao2021transform}, that requires $64$-point transforms.
In this paper, we extended the method proposed in~\cite{oliveira2019low} for larger blocklengths, i.e., for $N = 16$, $32$, and $64$.
To the best of our knowledge,
the current literature archives only
a few works~\cite{bas2010,bayer201216pt,thiago2017}
addressing low-complexity approximations for the DCT of the above-mentioned sizes.
We aim at proposing high-performing low-complexity DCT approximations for such blocklengths.

The paper is structured as follows.
In Section~\ref{s:hardware}, we present a brief review of the relevance of the DCT and approximations in the context of hardware implementation and its applications.
In Section~\ref{S:DCT}, we review the formulation of the DCT and low-complexity approximations.
Section~\ref{S:method} presents the methodology for
deriving the proposed transformations.
In Section~\ref{S:newappr}, the introduced DCT approximations are presented along with performance assessment and fast algorithms derivation.
Section~\ref{S:imagecomp} presents
computational experiments in image compression
that demonstrate the suitability of the suggested tools.
Section~\ref{S:conclusion} concludes the paper.

\section{Hardware review}\label{s:hardware}

The energy packing and high decorrelation properties
for
signals modeled after highly-correlated first-order Markovian process
makes the DCT
a widely employed method for image and video coding~\cite{rao1990}.
In this context,
the DCT
is
the main tool~\cite{9497745,canterle2020multiparametric,oliveira2019low,coelho2018efficient,8558684,9401178,9389991,7440320,7990539},
finding implementations in
JPEG~\cite{puchala2021approximate},
motion JPEG~\cite{9287147},
MPEG~\cite{9327927},
and
HEVC~\cite{9130767,8576601,7457337}.
Such efficient encoders
can provide a significant reduction in computational complexity,
enabling
the development of high-speed computing architecture~\cite{suresh2017approximate}.
Popular architectures that benefited from transform-based
encoding
include
HD and UHD videos~\cite{8954562,5159390},
smart antenna applications~\cite{thiripurasundari2018fpga,7042917},
secure image processing~\cite{rajapaksha2013asynchronous,madishetty2012vlsi},
image fusion~\cite{haghighat2011multi} and defusion~\cite{liang2015image},
biomedical signal processing~\cite{wahid2008efficient,wahid2011lossless},
to cite a few.

Therefore,
the
design of DCT-based very large-scale integration (VLSI) structures
has been an essential task for decreasing chip area, power, and time consumption~\cite{9301043,chung2021vlsi,1495723,8629129}.
Thus,
resource-constrained platforms
which
require
hardware designs capable of
larger autonomy, increased storage capacity, extended battery life,
and data transmission
are prime beneficiaries of
low-complexity methods.
This is illustrated
in the case of
low-powered devices in real-time applications~\cite{saponara2012real}
and nanotechnologies~\cite{9170577}, which require ultralow power consumption.
In addition, sensor architectures that
present stringent limitations on memory and processing speed~\cite{4912309,mechouek2016low,8938063} and
sub-branches like approximate memory,
which focus on the trade off between perfect data fidelity and storage density~\cite{ma2019approximate},
received technical contributions
from DCT approximations.
The requirements for low-cost computation
can be easily noticed in the context of
Internet of Things~(IoT)~\cite{8081677,hamza2019lightweight,7279738,margelis2019efficient} or
{5G} technologies~\cite{kansal2021efficient,potluri2012multiplier}, for example.
The assumption that approximate methods lead to hardware implementations of low resource consumption
was corroborated in a comprehensive study detailed in~\cite{puchala2021approximate}.

Motivated by this increasing demand,
we focus our work on further reducing the arithmetic complexity
of the DCT computation,
which directly affects hardware-oriented measures
such as
chip area,
dynamic power consumption,
critical path delay,
gate-count,
area-time,
and maximum clock frequency (throughput)~\cite{kulasekera2015multi,8890902,8558684,madanayake2020fast,jain2021fpga,9258670,9465468}.

\section{Exact and approximate DCT}
\label{S:DCT}

The $N$-point DCT is represented by an $N \times N$ matrix $\mathbf{C}_N$ whose elements are given by~\cite[pg.~61]{britanak2007discrete}:
\begin{equation}
	c_{i,j} = \sqrt{\frac{2}{N}} u_i \cos \left\{\frac{i(2j +1)\pi}{2N}\right\}, \quad i,j = 0,1, \ldots, N-1,
\end{equation}
where the quantities $u_i$ are defined as:
\begin{equation}
	u_i =
	\begin{cases}
		\frac{1}{\sqrt{2}}, &  \text{if $i = 0$,}\\
		1, &  \text{if $i \neq 0$.}
	\end{cases}
\end{equation}
Let $\mathbf{x} =
\begin{bmatrix}
	x_0 & x_1 & \ldots & x_{N-1}
\end{bmatrix}^\top$
be an $N$-point input vector.
The DCT transformation of $\mathbf{x}$ is the output vector
$\mathbf{X} =
\begin{bmatrix}
	X_0 & X_1 & \ldots & X_{N-1}
\end{bmatrix}^\top$, given by $\mathbf{X} = \mathbf{C}_N \cdot \mathbf{x}$. Since the DCT matrix is orthogonal, its inverse transformation can be written as $\mathbf{x} = \mathbf{C}_N^\top \cdot \mathbf{X}$~\cite[pg. 41]{britanak2007discrete}.

Generally, DCT approximations are transformations $\widehat{\mathbf{C}}_N$ that behave similarly to the exact DCT according to
a relevant criterion depending on the context at hand.
An approximate transform is usually based on
a low-complexity transform $\mathbf{T}_N$, i.e.,
a transformation matrix
whose
entries
possess very low multiplicative complexity~\cite{blahut2010,britanak2007discrete}.
Typical examples of low-complexity multipliers are:
$\{0, \pm 1, \pm 2, \pm 4, \ldots \}$,
$\{0,  \pm \tfrac{1}{2}, \pm 1, \pm 2, \pm 3 \}$,
and
$\{0, \pm \tfrac{1}{4}, \pm \tfrac{1}{2}, \pm 1\}$;
in all the above cases,
the multiplication of a given number by any set element requires no or a few additions and bit-shifting
operations.
Once a low-complexity matrix is obtained,
we can derive the associate approximate transform
according to the next formalism~\cite{cintra2014low,higham2008functions}:
\begin{align}
	\label{equation-dct-approximation}
	\widehat{\mathbf{C}}_N = \mathbf{S}_N \cdot \mathbf{T}_N,
\end{align}
where
\begin{align}
	\label{S_matrix}
	\mathbf{S}_N =	\sqrt{[\operatorname{diag}(\mathbf{T}_N\cdot\mathbf{T}_N^\top)]^{-1}},
\end{align}
being $\operatorname{diag}(\cdot)$ the diagonal matrix generated by its arguments
and $\sqrt{\cdot}$ the matrix square root operator~\cite{higham2008functions}.
If $\mathbf{T}_N$ is almost orthogonal~\cite{cintra2014low},
then the matrix $\widehat{\mathbf{C}}_N$ can represent a meaningful approximation for $\mathbf{C}_N$.
The concept of almost orthogonality stems from almost diagonality
property
as defined in~\cite{Flury1986,cintra2014low}.

Notice
that
if
$\mathbf{T}_N$
presents
the property of diagonality
($\mathbf{T}_N \cdot\mathbf{T}^\top_N=\left[\text{diagonal matrix}\right]$)
than
\eqref{equation-dct-approximation}
provides
an orthogonal approximation $\widehat{\mathbf{C}}_N$.
In addition,
if
$\mathbf{T}_N$
is
orthogonal
than
$\mathbf{S}_N$
results
in the identity matrix $\mathbf{I}_{N}$
and
$\widehat{\mathbf{C}}_N=\mathbf{T}_N$.

However,
finding a good low-complexity matrix~$\mathbf{T}_N$
can be a hard task,
because
of the large search space.
For instance,
the matrix space
of 16$\times$16
low-complexity matrices defined over
the set
$\{-1, 0, +1\}$
possesses
$3^{16^2} \approx 1.39\times 10^{122}$
candidate matrices.
An exhaustive search over this space
would take approximately
$4.41 \times 10^{105}$ years of computation
assuming that each matrix could be
generated and assessed in 1~nanosecond.

Therefore, a crucial step in deriving approximations
is the reduction of the search space
by restricting the search to potentially good matrices only.
Literature describes several methods to accomplish
such reduction:
(i)~matrix quantization~\cite{salomon2007};
(ii)~matrix parametrization~\cite{bas2011,ortega2004};
(iii)~algorithm parametrization~\cite{fw1992,tablada2015};
and
(iv)~visual inspection~\cite{Cintra2012,bas2008n,Bouguezel2008,bas2009,bas2010,haweel2016fast,Pati2010}.
Depending on the particular search space reduction approach,
the number of candidate approximations
can be
as small as just one matrix
(e.g., SDCT~\cite{Haweel2001} and RDCT~\cite{cintra2011dct})
or
as large as classes of various matrices,
as shown in~\cite{cintra2014low,oliveira2019low,potluri2014improved,tablada2015} for the
$8$-point case.

\section{DCT Approximations with minimal angular error}
\label{S:method}

In~\cite{oliveira2019low},
a search space reduction
based on the vector direction analysis~\cite{mardia2009,topics2001}
of the transformation basis vectors was proposed.
In the following,
we summarize the method.

The exact DCT matrix $\mathbf{C}_N$ can be understood as
a stack of row vectors $\mathbf{c}^{\top}_k$,
$k=0,1,\ldots,N-1$.
The goal of the method is to find
a low-complexity matrix $\mathbf{T}_N$,
whose rows are denoted by
$\mathbf{t}^{\top}_k$,
$k=0,1,\ldots,N-1$,
such that
a prescribed error measure
between the corresponding
rows of the exact DCT and the approximation
is minimized.
The entries of
$\mathbf{t}^{\top}_k$
are selected from
a set of
$\mathcal{D}= \{ d_0, d_1, \ldots, d_{D-1} \}$,
where
$d_i$,
$i=0,1,\ldots,D-1$,
are low-complexity (trivial~\cite{blahut2010}) multipliers.
Departing from the usual measures in the approximate transform literature,
such as
Euclidean distance,
in~\cite{oliveira2019low},
the angle between vectors is adopted as
the
error function.
Thus,
the following
optimization problem is defined~\cite{oliveira2019low}:
\begin{align} \label{eq:optim}
	\mathbf{t}_k
	=
	\underset
	{\mathbf{\mathbf{p}} \in \mathcal{D}^N}
	{\arg\min}
	\operatorname{angle}(\mathbf{p},\mathbf{c}_k)
	,
	\quad
	k=0,1,\ldots,N-1
	,
\end{align}
where
$\mathbf{p}^\top$ is a candidate row
defined over
the $N$-dimensional discrete space
$\mathcal{D}^N
=
\mathcal{D} \times \mathcal{D} \times \cdots \times \mathcal{D}$,
the angle between two vectors,
$\mathbf{p}$ and $\mathbf{c}_k$,
is given by
\begin{equation}
	\label{angle}
	\operatorname{angle}(\mathbf{p},\mathbf{c}_k)
	=
	\arccos
	\left(
	\frac
	{\langle\mathbf{p},\mathbf{c}_k\rangle}
	{||\mathbf{p}||\cdot||\mathbf{c}_k||}
	\right)
	,
\end{equation}
the symbol $\langle\cdot,\cdot\rangle$ denotes the usual inner product,
and
$||\cdot||$ is the norm induced by the inner product~\cite{Strang1988}.
The resulting low-complexity matrix~$\mathbf{T}_N$ from the above
optimization problem is said to have minimal angular error
relative to the exact DCT.

The search space implied by the above-described procedure
contains
$D^N$ rows.
Therefore,
the computation described in~\eqref{eq:optim}
requires
$N \cdot D^N$
angle evaluations at most.
In~\cite{oliveira2019low},
it was adopted
$\mathcal{D}=\{-1, 0, 1\}$ ($D=3$)
and
$N=8$,
implying
in only
$3^8 = 6561$ candidate rows.
For larger values of $N$,
the number of candidate rows
are
presented in Table~\ref{tab:Nvscandidates}, also
considering $\mathcal{D}=\{-1, 0, 1\}$.
Notice that any other discrete space with three elements
will generate the same number of candidate rows.
Hereafter
the collection of all possible rows
is
called $\mathcal{P}$.

\begin{table}[htb!]
	\centering

	\caption{The relation between matrix size and the number of candidate rows, considering a discrete space of three elements~(\textit{e.g.} $\mathcal{D}=\{-1, 0, 1\}$).}
	\label{tab:Nvscandidates}
	\begin{tabular}{ccc}
		\toprule
		$N$ &
		\begin{tabular}[c]{@{}c@{}}Number of  \\ candidate rows\end{tabular} \\
		\midrule
		8  & 6561                                                                              \\
		16  & 43046721	                                                                 \\
		32  & $\approx 1.85 \times10^{15} $                                                                 \\
		64  & $\approx 3.43 \times10^{30} $                                                                 \\
		\bottomrule
	\end{tabular}
\end{table}

The above procedure does not
ensure the orthogonality
of the resulting matrix.
Although orthogonality
is not strictly a necessary
condition for the derivation of good DCT approximations
(e.g.,~SDCT~\cite{Haweel2001}),
it is often a desirable design feature.
To address such specific need,
the optimization problem in~\eqref{eq:optim}
can
be extended
by the inclusion
of an orthogonality constraint
such that
each new candidate row
is
compared against the previously
obtained rows
ensuring that
their inner products are null~\cite{Seber2008}.
However,
the orthogonality constraint increases significantly the search time,
since this restriction is
sensitive to the order in which the rows
are approximated,
being feasible for small blocklengths only,
as successfully
demonstrated in~\cite{oliveira2019low}
for $N=8$.
Algorithm~\ref{alg:pseudo} presents
the pseudo-code that
contains the unconstrained procedure.

\begin{algorithm}[H]
	\caption{Pseudo-code for the unconstrained angle based method.}
	\label{alg:pseudo}
	\begin{algorithmic}[1]%
		\renewcommand{\algorithmicrequire}{\textbf{Input:}}
		\renewcommand{\algorithmicensure}{\textbf{Output:}}
		\REQUIRE $\mathbf{C}_N,\mathcal{P}$
		\ENSURE  $\mathbf{T}_N$
		\algrule
		\FOR {$k \leftarrow 0,1,\ldots,N-1$}
		\STATE $\theta_{\text{min}} \leftarrow2\pi$;
		\FOR {$i \leftarrow 0,1,\ldots,D^N-1$}
		\STATE $\mathbf{p} \leftarrow \mathcal{P}_i$;
		\STATE $\theta \leftarrow \operatorname{angle}(\mathbf{p},\mathbf{c}_k)$;
		\IF {($\theta < \theta_{\text{min}}$)}
		\STATE $\theta_{\text{min}} \leftarrow \theta$;
		\STATE $\mathbf{t}_k \leftarrow \mathbf{p}$;
		\ENDIF
		\ENDFOR
		\STATE $\mathbf{T}_N(k,:) \leftarrow \mathbf{t}_k$;
		\ENDFOR
		\RETURN $\mathbf{T}_N$;
	\end{algorithmic}
\end{algorithm}

\section{Proposed DCT approximations}\label{S:newappr}

In this section,
we report and assess
the proposed approximations
obtained from the approach based on the minimal angle error, presented in the Algorithm~\ref{alg:pseudo},
for
$N \in \{16, 32, 64\}$.
For this search, we have used a machine with the following specifications: hexa-core 4.5 GHz Intel(R) Core(R) I7-9750H, with 32 GB RAM running Ubuntu 20.04 LTS 64-bit and GPU GeForce RTX 2060.
In parallel, we also used a virtual machine from Google Cloud Platform with the following specifications: 8 cores 3.8 GHz Intel(Cascade Lake) with 32 GB RAM running Ubuntu 20.04 LTS 64-bit.
Besides the approximations obtained from this approach, we have also scaled the best transforms according to the JAM scaling method~\cite{Jridi2015} and
derived fast algorithms for the best performing
approximations.

Next, we present the considered design parameters to obtain the new approximate transforms,
their performance assessment and the proposed fast algorithms.
To the best of our knowledge, all of the obtained transforms are new in the lite\-ra\-tu\-re.

\subsection{Low-complexity matrices for	16-, 32-, and 64-point DCT approximations}

We solved the optimization problem
described in~\eqref{eq:optim}
considering the following parameters:
(i)~$N \in \{16, 32, 64\}$
and
(ii)
the following
sets of low-complexity multipliers:
$\mathcal{D}_1=\{0, \pm 1\}$,
$\mathcal{D}_2 =\{0, \pm \tfrac{1}{2}, \pm 1\}$,
$\mathcal{D}_3=\{0, \pm 1, \pm 2\}$,
$\mathcal{D}_4=\{0, \pm \tfrac{1}{4}, \pm \tfrac{1}{2}, \pm 1\}$,
$\mathcal{D}_5=\{0, \pm \tfrac{1}{2}, \pm 1, \pm 2\}$,
and
$\mathcal{D}_6=\{0, \pm \tfrac{1}{4}, \pm \tfrac{1}{2}, \pm 1, \pm 2\}$.
Such sets were separated because of the low-complexity nature of their elements.
In fact, multiplications by such elements require only bit-shifting operations.

The mapping from $\mathbf{T}_N$ to
$\hat{\mathbf{C}}_N$ shown in~\eqref{equation-dct-approximation}
is not one-to-one,
i.e.,
distinct low-complexity matrices
might result in the same approximation.
For instance,
the following 16-point low-complexity matrices,
${\mathbf{T}^{\prime}_{16}}$ and ${\mathbf{T}^{\prime\prime}_{16}}$,
lead to a single approximation:
\begin{equation*}
	\footnotesize
	{\mathbf{T}^{\prime}_{16}}=
	\left[\hspace{-0.1cm}\begin{rsmallmatrix}\\
		1 & 1 & 1 & 1 & 1 & 1 & 1 & 1 & 1 & 1 & 1 & 1 & 1 & 1 & 1 & 1 \\
		1 & 1 & 1 & 1 & 1 & 1 &   &   &   &   & -1 & -1 & -1 & -1 & -1 & -1 \\
		1 & 1 & 1 &   &   & -1 & -1 & -1 & -1 & -1 & -1 &   &   & 1 & 1 & 1 \\
		1 & 1 &   & -1 & -1 & -1 & -1 &   &   & 1 & 1 & 1 & 1 &   & -1 & -1 \\
		1 &   &   & -1 & -1 &   &   & 1 & 1 &   &   & -1 & -1 &   &   & 1 \\
		1 &   & -1 & -1 &   & 1 & 1 & 1 & -1 & -1 & -1 &   & 1 & 1 &   & -1 \\
		1 &   & -1 & -1 & 1 & 1 &   & -1 & -1 &   & 1 & 1 & -1 & -1 &   & 1 \\
		1 & -1 & -1 &   & 1 &   & -1 & -1 & 1 & 1 &   & -1 &   & 1 & 1 & -1 \\
		1 & -1 & -1 & 1 & 1 & -1 & -1 & 1 & 1 & -1 & -1 & 1 & 1 & -1 & -1 & 1 \\
		1 & -1 &   & 1 &   & -1 & 1 & 1 & -1 & -1 & 1 &   & -1 &   & 1 & -1 \\
		1 & -1 &   & 1 & -1 &   & 1 & -1 & -1 & 1 &   & -1 & 1 &   & -1 & 1 \\
		1 & -1 & 1 &   & -1 & 1 &   & -1 & 1 &   & -1 & 1 &   & -1 & 1 & -1 \\
		& -1 & 1 &   &   & 1 & -1 &   &   & -1 & 1 &   &   & 1 & -1 &   \\
		& -1 & 1 & -1 & 1 &   & -1 & 1 & -1 & 1 &   & -1 & 1 & -1 & 1 &   \\
		& -1 & 1 & -1 & 1 & -1 & 1 &   &   & 1 & -1 & 1 & -1 & 1 & -1 &   \\
		&   & 1 & -1 & 1 & -1 & 1 & -1 & 1 & -1 & 1 & -1 & 1 & -1 &   &   \\
	\end{rsmallmatrix}\hspace{-0.1cm}\right],
\end{equation*}
\begin{equation*}
	\footnotesize
	{\mathbf{T}^{\prime\prime}_{16}}=
	\left[\hspace{-0.1cm}\begin{rsmallmatrix}\\
		1 & 1 & 1 & 1 & 1 & 1 & 1 & 1 & 1 & 1 & 1 & 1 & 1 & 1 & 1 & 1 \\
		2 & 2 & 2 & 2 & 2 &   &   &   &   &   & -2 & -2 & -2 & -2 & -2 & -2 \\
		1 & 1 &   &   &   & -1 & -1 & -1 & -1 & -1 & -1 &   &   & 1 & 1 & 1 \\
		2 & 2 &   & -2 & -2 & -2 & -2 &   &   & 2 & 2 & 2 &   &   & -2 & -2 \\
		2 & 2 & -2 & -2 & -2 & -2 & 2 & 2 & 2 & 2 & -2 & -2 & -2 & -2 &   & 2 \\
		2 &   & -2 & -2 &   & 2 & 2 &   & -2 & -2 & -2 &   & 2 & 2 &   & -2 \\
		2 &   & -2 & -2 &   & 2 &   & -2 & -2 &   & 2 & 2 & -2 & -2 &   & 2 \\
		1 & -1 & -1 &   & 1 &   & -1 & -1 & 1 & 1 &   & -1 &   & 1 &   & -1 \\
		\frac{1}{2} & -\frac{1}{2} & -\frac{1}{2} & \frac{1}{2} & \frac{1}{2} & -\frac{1}{2} & -\frac{1}{2} & \frac{1}{2} & \frac{1}{2} & -\frac{1}{2} & -\frac{1}{2} & \frac{1}{2} &   & -\frac{1}{2} & -\frac{1}{2} & \frac{1}{2} \\
		2 & -2 &   & 2 &   & -2 &   & 2 & -2 & -2 & 2 &   & -2 &   & 2 & -2 \\
		& -2 &   & 2 & -2 &   & 2 & -2 & -2 & 2 &   & -2 & 2 &   & -2 & 2 \\
		& -2 & 2 &   & -2 & 2 &   & -2 & 2 &   & -2 & 2 &   & -2 & 2 & -2 \\
		2 & -2 & 2 & -2 & -2 & 2 & -2 & 2 & 2 & -2 & 2 & -2 & -2 & 2 & -2 &   \\
		& -\frac{1}{2} & \frac{1}{2} & -\frac{1}{2} &   &   & -\frac{1}{2} & \frac{1}{2} & -\frac{1}{2} & \frac{1}{2} &   & -\frac{1}{2} & \frac{1}{2} & -\frac{1}{2} & \frac{1}{2} &   \\
		& -2 & 2 & -2 & 2 & -2 &   &   &   & 2 & -2 & 2 & -2 & 2 & -2 &   \\
		&   &   & -2 & 2 & -2 & 2 & -2 & 2 & -2 & 2 & -2 & 2 & -2 &   &   \\
	\end{rsmallmatrix}\hspace{-0.1cm}\right],
\end{equation*}
since $\left(\sqrt{[\operatorname{diag}({\mathbf{T}^{\prime}_{16}}\cdot{\mathbf{T}^{\prime}_{16}}^\top)]^{-1}}\right)\cdot{\mathbf{T}^{\prime}_{16}}=\left(\sqrt{[\operatorname{diag}({\mathbf{T}^{\prime\prime}_{16}}\cdot{\mathbf{T}^{\prime\prime}_{16}}^\top)]^{-1}}\right)\cdot{\mathbf{T}^{\prime\prime}_{16}}$.
Such matrices
are referred to as equivalent~\cite{Seber2008}
and
can be grouped in equivalent classes.

Table~\ref{tab:generalResultsDCT_larger}
summarizes the number of matrices, equivalence classes
obtained
for 16-, 32-, and 64-point
low-complexity matrices,
and the sets considered for each blocklength. We do not consider all sets for the $32$- and $64$-point because of the computation time.
\begin{table}[htb!]
	\centering
	\caption{Total matrices and classes of equivalence obtained for the $16$-, $32$-, and $64$-point DCT}
	\label{tab:generalResultsDCT_larger}
	\begin{tabular}{cccc}
		\toprule
		\multicolumn{1}{c}{\begin{tabular}[c]{@{}c@{}}Sets\\  considered\end{tabular}} &
		$N$ &
		\begin{tabular}[c]{@{}c@{}}Number of matrices \\ obtained\end{tabular} &
		\begin{tabular}[c]{@{}c@{}}Number of classes \\ of equivalence\end{tabular} \\
		\midrule
		$\mathcal{D}_1$, $\mathcal{D}_2$, $\mathcal{D}_3$, $\mathcal{D}_4$, $\mathcal{D}_5$, $\mathcal{D}_6$ & 16  & 156                                                                    &       5                                                                      \\
		$\mathcal{D}_1$, $\mathcal{D}_2$, $\mathcal{D}_3$ & 32  & 3                                                                      &       2                                                                     \\
		$\mathcal{D}_1$ & 64  & 1                                                                      &       1                                                                     \\
		\bottomrule
	\end{tabular}
\end{table}
The 16-, 32-, and 64-point low-complexity matrices
obtained
are denoted here as $\mathbf{T}_{N, i}$,
where $i$ and $N$ refers to the equivalence class
and the length of the low-complexity matrix, respectively.
For the case $N =$ 16,
we have five new approximations obtained directly from
the proposed method;
for  $N = $ 32,
two new matrices;
and, considering $N =$ 64,
we have only one new matrix.
The best performing matrices (as discussed in Section~\ref{ss:performaceassess}) are numerically shown in the Appendix.

Due to the size of the search space for some settings,
we had to reduce the number of matrices according to the following procedure:
\begin{enumerate}
	\item
	Split the exact DCT into two matrices as follows:
	\begin{equation*}
		\mathbf{C}_N = \operatorname{abs}(\mathbf{C}_N) \circ \operatorname{sign}(\mathbf{C}_N)
		,
	\end{equation*}
	where $\operatorname{abs}(\cdot)$~\cite[pg.~10]{bartle2000introduction}
	returns
	the absolute value of its input,
	$\operatorname{sign}(\cdot)$
	is
	the
	entry-wise signum function~\cite[pg.~286]{britanak2007discrete},
	and $\circ$
	represents
	the element wise multiplication~\cite[pg.~251]{Seber2008};

	\item
	Select $\mathcal{D}$ and remove its negative elements, \textit{e.g.}: $\mathcal{D}^+_3 =\left\{0,1,2\right\}$;

	\item
	Rewrite~\eqref{eq:optim} replacing $\mathbf{c}_k$ and $\mathcal{D}$ with $\operatorname{abs}(\mathbf{c}_k)$ and $\mathcal{D}^+$, respectively, as follows:
		\begin{align} \label{eq:optim2}
			\mathbf{t}^*_k
			=
			\underset
			{\mathbf{\mathbf{p}} \in (\mathcal{D}^+)^N}
			{\arg\min}
			\operatorname{angle}(\mathbf{p},\operatorname{abs}(\mathbf{c}_k))
			,
			\quad
			k=0,1,\ldots,N-1
			;
		\end{align}

	\item
	Use~\eqref{eq:optim2} to approximate the rows of $\operatorname{abs}(\mathbf{C}_N)$ individually and obtain the low-complexity matrix $\mathbf{{T}}^*_N$;

	\item
	Obtain the low-complexity $\mathbf{{T}}_N$ by:
	\begin{align*}
		\mathbf{{T}}_N=\mathbf{{T}}^*_N \circ \operatorname{sign}(\mathbf{C}_N)
		.
	\end{align*}
\end{enumerate}

To further reduce the search space,
the symmetries of $\operatorname{abs}(\mathbf{C}_N)$ were used.
Notice that the $j$th and the $(N-j-1)$th column of $\operatorname{abs}(\mathbf{C}_N)$
are
equals.
Thus,
we only need to find approximations for half of the values in each row.
Using this property in~\eqref{eq:optim2},
the computation
requires
$N \cdot (D^+)^{\frac{N}{2}}$
angle evaluations at most.
This search space reduction
was only adopted
for $N=32$ considering $\mathcal{D}_2$ and $\mathcal{D}_3$;
and
for $N=64$ considering $\mathcal{D}_1$.

\subsection{JAM scaling method}

In addition to the obtained matrices,
we also derive
scaled approximations
according to the JAM scaling method described in~\cite{Jridi2015}.
A scaled approximation derived from the JAM method
consists of an $2N$-point transformation
based on an $N$-point approximation
related as follows:
\begin{equation}
	\label{JAM_scaling}
	\mathbf{T}_{2N} =
	\begin{bmatrix}
		\mathbf{P}_1 & \mathbf{P}_2
	\end{bmatrix}
	\cdot
	\begin{bmatrix}
		\mathbf{T}_{N} & \\
		& \mathbf{T}_{N}
	\end{bmatrix}
	\cdot
	\begin{bmatrix}
		\mathbf{I}_{N} & \bar{\mathbf{I}}_{N} \\
		\bar{\mathbf{I}}_{N}  & \mathbf{I}_{N}
	\end{bmatrix},
\end{equation}
where $\mathbf{T}_{N}$ is an  $N$-point approximation, $\mathbf{I}_{N}$ and $\bar{\mathbf{I}}_{N}$ are, respectively, the $N$-point identity and counter-identity matrices. Matrices $\mathbf{P}_1$  and $\mathbf{P}_2$ are permutations of $2N \times N$ dimension. Matrix $\mathbf{P}_1$  contains ones in positions $(2i,i),i=0,\ldots,N-1$, and zeros elsewhere; whereas $\mathbf{P}_2$ presents ones in positions $(2i+1,i),i=0,\ldots,N-1$, and zeros elsewhere.

Thus,
we scaled the low-complexity matrices that generate the
best performing transforms for $N=$16 and 32,
according to the assessment measurements displayed in Table~\ref{tab:apps_measures},
through the JAM method.
The scaling method is a function $f$ defined according to:
\begin{eqnarray*}
	f: &\mathbb{R}^{N \times N}& \longmapsto  \mathbb{R}^{2N \times 2N} \\
	&\mathbf{T}_N& \longmapsto \mathbf{T}_{2N} = f(\mathbf{T}_N).
\end{eqnarray*}
We denote $\mathbf{T}_N^{(j)}$ as the $2^j \cdot N \times 2^j \cdot N$ matrix based on $\mathbf{T}_N$:
\begin{equation*}
	\mathbf{T}_N^{(j)} \triangleq f(f(\ldots f(\mathbf{T}_N))),
\end{equation*}
where  function $f$ is applied $j$ times.

The proposed method in~\cite{Jridi2015}
has
a term of $\frac{1}{\sqrt{2}}$ that multiplies \eqref{JAM_scaling}
which
was previously omitted in the scaling to be integrated in the final quantization. Thus, the approximation $\mathbf{\widehat{C}}_{N}^{(j)}$
is
obtained by
\begin{equation}
		\mathbf{\widehat{C}}_{N}^{(j)} = \left(\frac{1}{\sqrt{2}}\right)^j \cdot \mathbf{S}^*_N \cdot \mathbf{T}_N^{(j)},
\end{equation}
where $\mathbf{S}^*_N$ is computed from $\mathbf{T}_N^{(j)}$ as detailed in~\eqref{S_matrix}, \textit{mutatis mutandis}. For instance, $\mathbf{\widehat{C}}_{16}^{(2)}$ is the result of two JAM applications in a 16-point matrix, resulting in a 64-point matrix.

\subsection{Performance assessment}\label{ss:performaceassess}

We evaluated the proposed approximations
according to
similarity measures and coding measures.
For similarity measures,
we considered
the mean square error
($\operatorname{MSE}$)~\cite{britanak2007discrete,ochoa2019discrete}
and
the total energy error ($\epsilon$)~\cite{cintra2011dct};
whereas for coding measures,
we adopted
the unified coding gain ($C_g$)~\cite{Yasuda1991cg}
and
the transform efficiency ($\eta$)~\cite{britanak2007discrete}.
Based on the deviation from diagonality~\cite{Flury1986},
we can also quantify the deviation from orthogonality ($\delta(\cdot)$)
of
the discussed approximations;
this measure
informs how close to orthogonality a matrix is and is given by:
\begin{equation}
	\delta(\mathbf{A}) = 1- \frac{|| \operatorname{diag}(\mathbf{A})||_F}{||\mathbf{A}||_F},
\end{equation}
where $\mathbf{A}$ is a square matrix and $||\cdot||_F$ is the Frobenius norm for matrices~\cite{Seber2008}.

For comparison purposes,
we considered a comprehensive set of
approximations archived in literature.
For $N=16$,
the following approximations
were separated:
(i)~the approximation proposed by Haweel~($\mathbf{\widehat{C}}_{16,\text{SDCT}}$)~\cite{Haweel2001}
(ii)~the approximation proposed by
Jridi, Alfalou, and Meher~($\mathbf{\widehat{C}}_{16,\text{JAM}}$)~\cite{Jridi2015};
(iii)~the approximation proposed by Bouguezel, Ahmad, and Swamy, ($\mathbf{\widehat{C}}_{16,\text{BAS}}$),
in~\cite{bas2010};
(iv)~the approximation proposed in~\cite{bayer201216pt},
($\mathbf{\widehat{C}}_{16,\text{BCEM}}$);
(v)~the approximation introduced in~\cite{thiago2017}, ($\mathbf{\widehat{C}}_{16,\text{SOBCM}}$), and
(vi)~the scaled approximation proposed in~\cite{oliveira2019low} ($T_{(16)}$), ($\mathbf{\widehat{C}}_{16,\text{OCBSML}}$).

For $N=32$, we considered the following methods:
(i)~the approximation proposed by Haweel~($\mathbf{\widehat{C}}_{32,\text{SDCT}}$)~\cite{Haweel2001};
(ii)~the approximation proposed by
Jridi, Alfalou, and Meher~($\mathbf{\widehat{C}}_{32,\text{JAM}}$)~\cite{Jridi2015};
(iii)~the approximation proposed by Bouguezel, Ahmad, and Swamy, ($\mathbf{\widehat{C}}_{32,\text{BAS}}$),
in~\cite{bas2010};
(vi)~the scaled approximation proposed in~\cite{oliveira2019low} ($T_{(32)}$), ($\mathbf{\widehat{C}}_{32,\text{OCBSML}}$).

For $N=64$, we considered the following approximations:
(i)~the approximation proposed by Haweel~($\mathbf{\widehat{C}}_{64,\text{SDCT}}$)~\cite{Haweel2001};
(ii)~the scaled version of the approximation proposed in~\cite{oliveira2019low}, ($\mathbf{\widehat{C}}_{64,\text{OCBSML}}$).

Table~\ref{tab:apps_measures}
displays the results for the performance measures
of the proposed approximations
along with the competing approximations and the exact $N$-point DCT ($\mathbf{C}_N$).
For each value of $N$, we highlighted the best measurements of each metric.
\begin{table}[htb!]
	\centering
	\caption{Performance measures for the DCT approximations in literature and the
		new approximations proposed }
	\label{tab:apps_measures}
	\begin{tabular}{lccccc}
		\toprule
		Approximation
		& $\epsilon(\widehat{\mathbf{C}})$ & $\operatorname{MSE}(\widehat{\mathbf{C}})$
		& $C_g(\widehat{\mathbf{C}})$ & $\eta(\widehat{\mathbf{C}})$ & $\delta(\widehat{\mathbf{C}})$\\
		\midrule
		\multicolumn{6}{c}{$N=16$}\\ \midrule
		$\mathbf{C}_{16}$                  &	0	&	0	&	 9.4555 &	 88.4518 &0 \\
		\hdashline
		$\mathbf{\widehat{C}}_{\text{16},1}$ 	&	 3.7043 &	 0.0172 &	 7.7474 &	 70.5034& 0.0423\\
		$\mathbf{\widehat{C}}_{\text{16},2}$ 	&	 3.7043 &	 0.0172 &	 8.2190 &	 70.6902& 0.0136\\
		$\mathbf{\widehat{C}}_{\text{16},3}$ 	&	 1.0227 &	 0.0054 &	 8.9653 &	 78.4016& 0.0239\\
		$\mathbf{\widehat{C}}_{\text{16},4}$ 	&	 0.6337 &	 0.0035 &	 9.0922 &	 80.1145& 0.0118 \\
		$\mathbf{\widehat{C}}_{\text{16},5}$ 	&	 \textbf{0.5748} &	 \textbf{0.0031} &	 \textbf{9.1268} &	 \textbf{80.4401} &0.0060\\
		$\mathbf{\widehat{C}}_{16,\text{SDCT}}$ & 8.2537  &   0.0429 & 6.0297  & 64.9653 & 0.1056\\
		$\mathbf{\widehat{C}}_{16,\text{JAM}}$ & 14.7402 & 0.0506 & 8.4285 & 72.2296 & 0\\
		$\mathbf{\widehat{C}}_{16,\text{BAS}}$ & 16.4071 & 0.0564 & 8.5208 & 73.6345 & 0 \\
		$\mathbf{\widehat{C}}_{16,\text{BCEM}}$ & 8.0806  & 0.0465 & 7.8401 & 65.2789& 0\\
		$\mathbf{\widehat{C}}_{16,\text{SOBCM}}$  & 40.9996 & 0.0947 & 7.8573 &  67.6078 & 0\\

		$\mathbf{\widehat{C}}_{16,\text{OCBSML}}$	&	13.7032 &	0.0474  &	8.8787  &	76.8108&  0\\
		\midrule
		\multicolumn{6}{c}{$N=32$}\\ \midrule
		$\mathbf{C}_{32}$           &	0	&	0	&	 9.7736 &	 81.6962  &0\\
		\hdashline
		$\mathbf{\widehat{C}}_{\text{32},1}$ 	&	 7.6403 &	 0.0287 &	 7.4624 &	 52.5455& 0.0586\\
		$\mathbf{\widehat{C}}_{\text{32},2}$ 	&	\textbf{ 2.3525 }&	 \textbf{0.0100} &	9.0983 &	 64.9265& 0.0190\\
		$\mathbf{\widehat{C}}_{16,5}^{(1)}$ 	& 30.0539	&	  0.0829 &	\textbf{9.1939 } &	 \textbf{64.9983 }& 0.0059 \\
		$\mathbf{\widehat{C}}_{32,\text{SDCT}}$ & 18.2386   &   0.0748   & 5.5623   & 41.6653  & 0.1472 \\
		$\mathbf{\widehat{C}}_{32,\text{JAM}}$ & 48.0956 & 0.1124 & 8.5010 & 56.9700 & 0\\
		$\mathbf{\widehat{C}}_{32,\text{BAS}}$ &  57.1260 & 0.1171 & 8.4971 & 58.1727& 0\\

		$\mathbf{\widehat{C}}_{32,\text{OCBSML}}$	&	46.2658 &	0.1104  &	8.9505  &	61.0272  &0\\
		\midrule
		\multicolumn{6}{c}{$N=64$}\\ \toprule
		$\mathbf{C}_{64}$                &	0	&	0	&	 9.9366 &	 75.55406 &0\\
		\hdashline
		$\mathbf{\widehat{C}}_{\text{64},1}$ 	&	\textbf{15.5707}	&	 \textbf{0.0434} &	 7.2436 &	 36.4275&  0.0594\\
		$\mathbf{\widehat{C}}_{16,5}^{(2)}$ 	&  103.2435&0.1833&\textbf{9.2144}&\textbf{51.6925}&0.0059\\
		$\mathbf{\widehat{C}}_{32,2}^{(1)}$ 	&  66.8310	&  0.1355 &	 9.1164 & 51.2582& 0.0190 \\
		$\mathbf{\widehat{C}}_{64,\text{SDCT}}$ & 38.2630   & 0.1141   & 5.2192   &  27.9725  & 0.1520  \\
		$\mathbf{\widehat{C}}_{64,\text{OCBSML}}$&	125.2247 &	0.2015  &	8.9748 &	48.4443 &0\\
		\bottomrule
	\end{tabular}
\end{table}
We identify $\mathbf{\widehat{C}}_{\text{16},5}$, $\mathbf{\widehat{C}}_{\text{32},2}$, $\mathbf{\widehat{C}}_{16,5}^{(1)}$, $\mathbf{\widehat{C}}_{\text{64},1}$, and $\mathbf{\widehat{C}}_{16,5}^{(2)}$  as the best transforms for $N=16$, $32$, and $64$, respectively.
Approximations $\mathbf{\widehat{C}}_{\text{16},5}$, $\mathbf{\widehat{C}}_{\text{32},2}$, and $\mathbf{\widehat{C}}_{\text{64},1}$
are the low-complexity matrices obtained from the the fifth, second, and first class of equivalence for $N=16$, $32$, and $64$, respectively. The approximations $\mathbf{\widehat{C}}_{16,5}^{(1)}$ and $\mathbf{\widehat{C}}_{16,5}^{(2)}$ are the scaled approximations based on $\mathbf{\widehat{C}}_{\text{16},5}$.
These proposed approximations outperformed the DCT approximations already known in the literature.

\subsection{Fast algorithms}\label{S:FastAlgo}

In order to reduce the arithmetic cost we factorized the best transforms into sparse matrices considering usual decimation-based techniques~\cite[pg.~74]{blahut2010}.
The factorization for the proposed transforms was developed using butterfly-based structures, such as in~\cite{hou87,yip1988,rao1990,cintra2011dct,cintra2014low}.
The complexity of the fast algorithms was evaluated in terms of the number of arithmetic operations.  The arithmetic complexity does not depend on the available technology,
an issue that occurs in measures such as computation time~\cite{oppenheim1999discrete,levitin2008introduction}. Table~\ref{tab:apps_overview} presents the arithmetic complexity of the proposed transforms before and after the matrix factorization and also the percentage of complexity reduction.

\begin{table}[htb!]
	\centering
	\footnotesize
	\caption{The arithmetic complexity of the proposed transforms before and after the matrix factorization.}
	\label{tab:apps_overview}
	\begin{tabular}{c@{\hspace{-1ex}}ccccccccc}
		\toprule
		&\multicolumn{3}{c}{Before factorization}&\multicolumn{2}{c}{After factorization}&\multicolumn{2}{c}{Reduction (\%)}\\
		\multicolumn{1}{c}{\begin{tabular}[c]{@{}c@{}}Matrix \end{tabular}} &
		Mult& Adds & Bit-shifting & Adds & Bit-shifting& Adds & Bit-shifting\\
		\midrule
		\multicolumn{8}{c}{$N=16$}\\ \midrule
		$\mathbf{C}_{16}$	&256&	240 	&	0&-&-&-&-\\
		\hdashline
		$\mathbf{T}_{\text{16},1}$ &0	&	184	&	0 &82& 0 &55.44&0\\             		
		$\mathbf{T}_{\text{16},2}$ &0	&	192	&	0  &80&0 &58.33&0\\						
		$\mathbf{T}_{\text{16},3}$ &0	&	208	&	80  &88&30&57.69&62.50\\				
		$\mathbf{T}_{\text{16},4}$ &0&	224 	&	96 &92&34&58.93&73.44\\				
		$\mathbf{T}_{\text{16},5}$ &0		&	240 	&	160 &100&62&58.33&64.77\\
		$\mathbf{{T}}_{16,\text{OCBSML}}$&0&208&96&64&12&69.23&79.17\\
		\midrule
		\multicolumn{8}{c}{$N=32$}\\ \midrule
		$\mathbf{C}_{32}$ &1024	&	992	&	0&-&-&-&-\\
		\hdashline
		$\mathbf{T}_{\text{32},1}$ 	&0	&	752	&0 &287&0&61.83&0\\
		$\mathbf{T}_{\text{32},2}$ 	&0&	864	& 320& 328&110&62.04&65.62	\\
		$\mathbf{T}_{16,5}^{(1)}$ &0&992&640&232&124&76.61&80.62\\
		$\mathbf{{T}}_{32,\text{OCBSML}}$ &0&864&384&160&24&81.48&89.58\\
		\midrule
		\multicolumn{8}{c}{$N=64$}\\ \midrule
		$\mathbf{C}_{64}$	&4096&	4032	&	0&-&-&-&-\\
		\hdashline
		$\mathbf{T}_{\text{64},1}$ 	&0	&	3040	&	0&1087&0 &64.24&0\\
		$\mathbf{T}_{16,5}^{(2)}$ &0&4032&2560&528&248&86.90&90.31\\
		$\mathbf{T}_{32,2}^{(1)}$ &0&3520&1280&720&220&79.54&82.81\\
		$\mathbf{{T}}_{64,\text{OCBSML}}$ &0&3520&1536&384&48&89.09&94.79\\
		\bottomrule
	\end{tabular}
\end{table}

Although
the matrices
$\mathbf{{T}}_{16,\text{OCBSML}},\mathbf{{T}}_{32,\text{OCBSML}}$, and $\mathbf{{T}}_{64,\text{OCBSML}}$~\cite{oliveira2019low} are more benefited by factorization,
with a few more operations, the proposed approximations can be used, providing better performance.
According to Table~\ref{tab:apps_measures} and~\ref{tab:apps_overview},
when comparing
$\mathbf{{T}}_{16,\text{OCBSML}}$ and $\mathbf{T}_{16,5}$,
the proposed approximation
requires
only 16 more additions
but
it
presents
a reduction of approximately 96\% and 93\%
in terms of energy error and MSE, respectively;
and
a gain of
approximately 2.8\% in coding gain and 4.7\% in transform efficiency.

Considering $N=32$,
the proposed approximation $\mathbf{T}_{32,1}$
needs $127$ extra additions than $\mathbf{{T}}_{32,\text{OCBSML}}$
and
it
presents a reduction of approximately $95\%$ and $93\%$ of energy error and MSE, respectively;
and
a improvement of $1.6\%$ in coding gain and $6.4\%$ in transform efficiency.
In addition,
the proposed approximation $\mathbf{T}^{(1)}_{16,5}$
requires
$72$ more additions than $\mathbf{{T}}_{32,\text{OCBSML}}$
and
it
presents
a reduction of $54\%$ and $33\%$ of energy error and MSE, respectively;
and
a gain of
approximately $2.7\%$ in coding gain and $6.5\%$ in transform efficiency.
Approximations for $N=64$ (except the SDCT)
were
first presented in this paper.

For better understanding and reproducibility,
each sparse matrix used
to obtain low-complexity matrices
associated with the optimal transforms $\mathbf{T}_{16,5}$, $\mathbf{T}_{32,2}$, and $\mathbf{T}_{64,1}$ are detailed in Appendix.

\section{Image compression experiments}\label{S:imagecomp}

To evaluate the performance
of the proposed 16-, 32-, and 64-point DCT
approximations we performed JPEG-like image compression experiments, as in~\cite{cintra2011dct,cintra2014low,potluri2012multiplier}. We considered $45$ $8$-bit images of $512\times512$ obtained from~\cite{uscsipi}. All images were subdivided into $N \times N$ sub-blocks and were submitted to the following 2D transformation.

Let $\mathbf{A}$ be an $N\times N$ sub-block. Two
approaches will be considered to compute the inverse transformation: one using the inverse of the approximation, and another one using its transpose.
Thus,
the direct and inverse transformations in the first approach, referred to as Method I, induced by $\widehat{\mathbf{C}}_N$ are given, respectively, by~\cite{suzuki2010integer, cintra2014low, cintra2011dct}:
\begin{align} \label{eq:AB}
	\mathbf{B} = \widehat{\mathbf{C}}_N \cdot \mathbf{A} \cdot \widehat{\mathbf{C}}_N^{\top} \quad \text{and} \quad
	\mathbf{A} = \widehat{\mathbf{C}}_N^{-1} \cdot \mathbf{B} \cdot ({\widehat{\mathbf{C}}_N^{-1}})^{\top},
\end{align}
where $\mathbf{A}$ and $\mathbf{B}$ are $N \times N$ matrices.
In the second approach,
referred to as Method II,
the direct and inverse transformations
are given, respectively, by:
\begin{align} \label{eq:AB}
	\mathbf{B} = \widehat{\mathbf{C}}_N \cdot \mathbf{A} \cdot \widehat{\mathbf{C}}_N^{\top} \quad \text{and} \quad
	\mathbf{A} = \widehat{\mathbf{C}}_N^{\top} \cdot \mathbf{B} \cdot \widehat{\mathbf{C}}_N
	.
\end{align}

To evaluate the exact DCT,
the transformation matrix $\widehat{\mathbf{C}}_N$ and $\widehat{\mathbf{C}}_N^{-1}$ (or $\widehat{\mathbf{C}}_N^{\top}$ according to the approach)
are
replaced by $\mathbf{C}_N$ and $\mathbf{C}_N^{-1}$, respectively.
Considering the zig-zag pattern~\cite[pg.~30]{salomon2007}, we retained the initial $r$ coefficients from each sub-block $\mathbf{B}$. Finally, we applied the inverse 2D transform in each sub-block, and then the compression images are obtained.

Original and compressed images were evaluated considering usual quality assessment measures:
(i)~the mean square error (MSE)~\cite{britanak2007discrete},
(ii)~the peak signal-to-noise ratio (PSNR)~\cite{salomon2007},
and (iii)~the mean structural similarity index (MSSIM)
\cite{Wang2004}.
Notice that the MSE measures are computed considering the original and
the compressed images; not to be confused with the MSE calculation
described in Section~\ref{ss:performaceassess}.
Although the MSE and PSNR measures are popular in the context of image compression, it was shown in~\cite{Wang2009mse} that they might offer limited results as image quality tools. On the other hand, the MSSIM was shown to be a better measure when it comes to capturing the image quality as understood by the human visual system model~\cite{Wang2004,Wang2009mse}.

The image compression experiments were divided into two steps: (i)~a qualitative analysis where we considered the compressed \textit{Peppers} image with a compression rate of approximately $80\%$; (ii)~and a quantitative analysis, where we considered the average measurements from $45$ standardized images.
Both analysis are presented next.

\subsection{Qualitative analysis}
\begin{figure*}[h!]
	\centering
	\subfloat[$\mathbf{C}_{16}$]{\includegraphics[scale = 0.625]{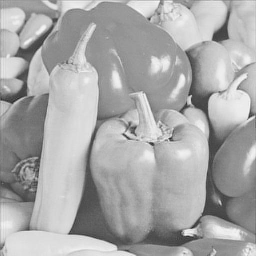}} \hspace{1mm}
	\subfloat[$\mathbf{\widehat{C}}_{16,\text{SDCT}}$]{\includegraphics[scale = 0.625]{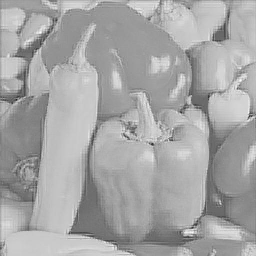}} \hspace{1mm}
	\subfloat[$\mathbf{\widehat{C}}_{16,\text{BAS}}$]{\includegraphics[scale = 0.625]{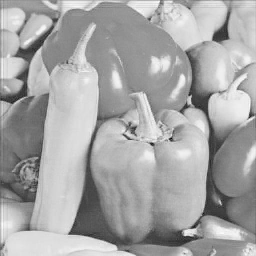}} \\
	\subfloat[$\mathbf{\widehat{C}}_{16,\text{JAM}}$]{\includegraphics[scale = 0.625]{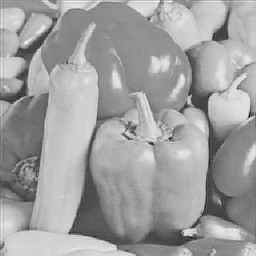}}\hspace{1mm}
	\subfloat[$\mathbf{\widehat{C}}_{16,\text{OCBSML}}$]{\includegraphics[scale = 0.625]{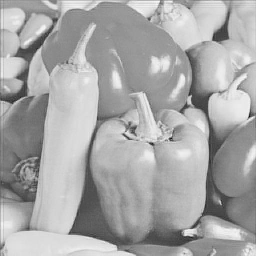}}\hspace{1mm}
	\subfloat[$\mathbf{\widehat{C}}_{\text{16},5}$]{\includegraphics[scale = 0.625]{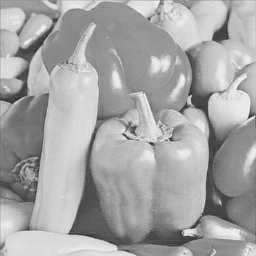}}
	\caption{Compressed \textit{Peppers} image for $N=16$ considering $r=50$ considering Method I.}
	\label{fig:pep-16}
\end{figure*}

\begin{figure*}[h!]
	\centering
	\subfloat[$\mathbf{C}_{32}$]{\includegraphics[scale = 0.625]{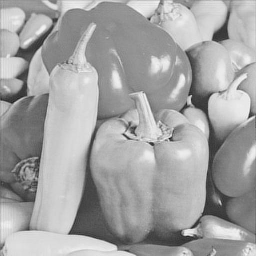}} \hspace{1mm}
	\subfloat[$\mathbf{\widehat{C}}_{32,\text{SDCT}}$]{\includegraphics[scale = 0.625]{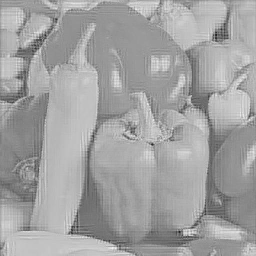}} \hspace{1mm}
	\subfloat[$\mathbf{\widehat{C}}_{32,\text{BAS}}$]{\includegraphics[scale = 0.625]{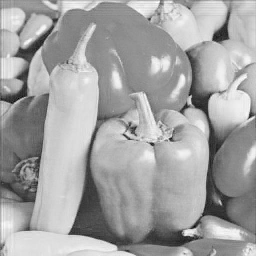}} \hspace{1mm}
	\subfloat[$\mathbf{\widehat{C}}_{32,\text{JAM}}$]{\includegraphics[scale = 0.625]{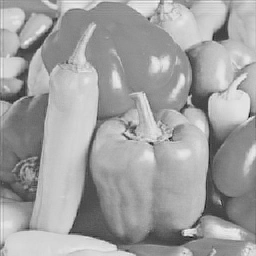}} \hspace{1mm}
	\subfloat[$\mathbf{\widehat{C}}_{32,\text{OCBSML}}$]{\includegraphics[scale = 0.625]{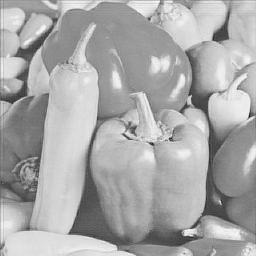}}\hspace{1mm}
	\subfloat[$\mathbf{\widehat{C}}_{\text{32},2}$]{\includegraphics[scale = 0.625]{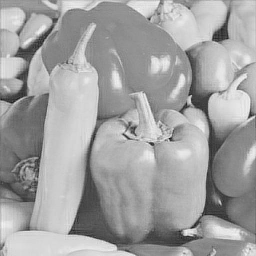}} \hspace{1mm}
	\subfloat[$\mathbf{\widehat{C}}_{16,5}^{(1)}$]{\includegraphics[scale = 0.625]{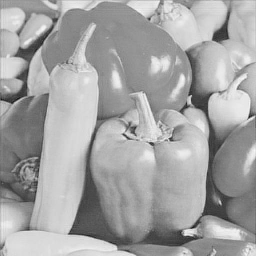}}
	\caption{Compressed \textit{Peppers} image for $N=32$ considering $r=205$ considering Method I.}
	\label{fig:pep-32}
\end{figure*}

\begin{figure*}[h!]
	\centering
	\subfloat[$\mathbf{C}_{64}$]{\includegraphics[scale = 0.625]{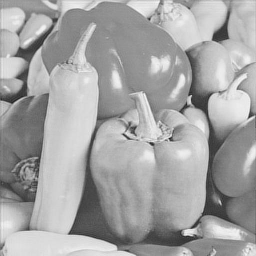}} \hspace{1mm}
	\subfloat[$\mathbf{\widehat{C}}_{64,\text{SDCT}}$]{\includegraphics[scale = 0.625]{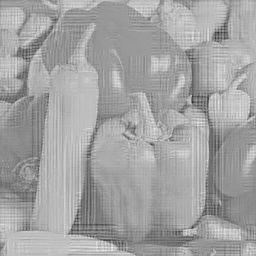}} \hspace{1mm}
	\subfloat[$\mathbf{\widehat{C}}_{64,\text{OCBSML}}$]{\includegraphics[scale = 0.625]{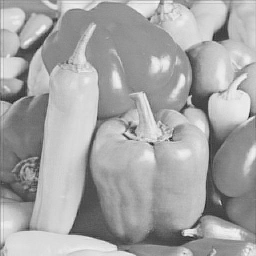}}\hspace{1mm}\\
	\subfloat[$\mathbf{\widehat{C}}_{\text{64},1}$]{\includegraphics[scale = 0.625]{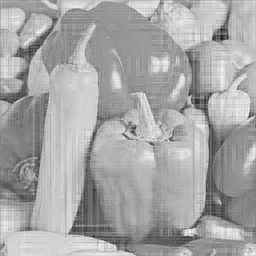}} \hspace{1mm}
	\subfloat[$\mathbf{\widehat{C}}_{16,5}^{(2)}$]{\includegraphics[scale = 0.625]{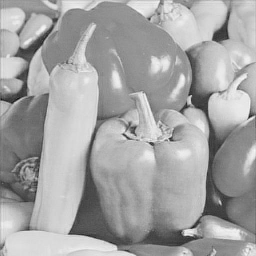}}\hspace{1mm}
	\subfloat[$\mathbf{\widehat{C}}_{32,2}^{(1)}$]{\includegraphics[scale = 0.625]{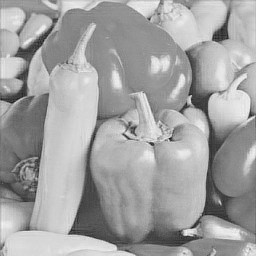}}
	\caption{Compressed \textit{Peppers} image for $N=64$ considering $r=820$ considering Method I.}
	\label{fig:pep-64}
\end{figure*}

For the qualitative analysis, we considered the compressed \textit{Peppers} image~\cite{uscsipi},
with a compression rate (CR) of approximately $80\%$. The compression rate is defined by:
\begin{eqnarray*}
	\text{CR} = 1 - \frac{r}{N^2}.
\end{eqnarray*}
Fig.~\ref{fig:pep-16}, \ref{fig:pep-32}, and \ref{fig:pep-64} display the compressed \textit{Peppers} images with the DCT and approximations, for $N=16$, $32$, and $64$, respectively.
Visually, the reconstructed images after the compression with the proposed transforms exhibit quality comparable to the ones compressed using the exact DCT and known approximations. For a better representation, we present
the MSE, PSNR, and MSSIM
of these images
in Table~\ref{tab:image-measures}.
The best results for each sample image and measure
are highlighted.
\textbf{The proposed approximations
	for the 16-, 32-, and  64-point DCT
	show better results than
	the approximations in literature.}
To the best of our knowledge, there is no
64-point DCT approximation in the literature for comparison, except for the $\widehat{\mathbf{{C}}}_{64,\text{SDCT}}$~\cite{Haweel2001}.

\begin{figure*}[h!]
	\centering
	\subfloat[$\mathbf{C}_{16}$]{\includegraphics[scale = 0.625]{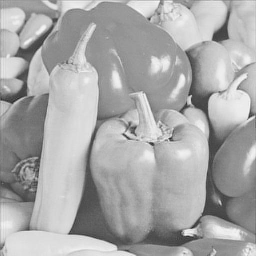}} \hspace{1mm}
	\subfloat[$\mathbf{\widehat{C}}_{16,\text{SDCT}}$]{\includegraphics[scale = 0.625]{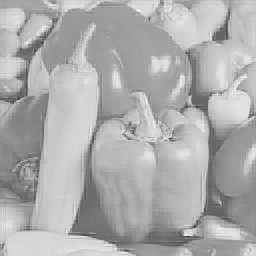}} \hspace{1mm}
	\subfloat[$\mathbf{\widehat{C}}_{16,\text{BAS}}$]{\includegraphics[scale = 0.625]{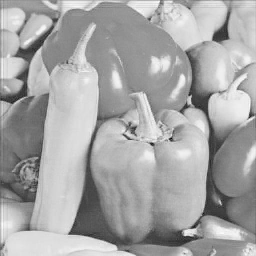}} \\
	\subfloat[$\mathbf{\widehat{C}}_{16,\text{JAM}}$]{\includegraphics[scale = 0.625]{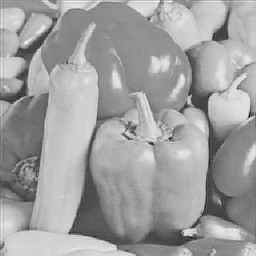}}\hspace{1mm}
	\subfloat[$\mathbf{\widehat{C}}_{16,\text{OCBSML}}$]{\includegraphics[scale = 0.625]{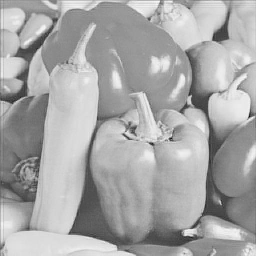}}\hspace{1mm}
	\subfloat[$\mathbf{\widehat{C}}_{\text{16},5}$]{\includegraphics[scale = 0.625]{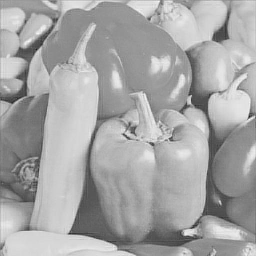}}
	\caption{Compressed \textit{Peppers} image for $N=16$ considering $r=50$ considering Method II.}
	\label{fig:pep-16_methodII}
\end{figure*}

\begin{figure*}[h!]
	\centering
	\subfloat[$\mathbf{C}_{32}$]{\includegraphics[scale = 0.625]{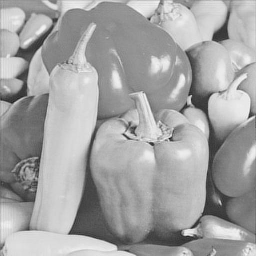}} \hspace{1mm}
	\subfloat[$\mathbf{\widehat{C}}_{32,\text{SDCT}}$]{\includegraphics[scale = 0.625]{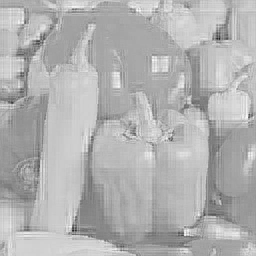}} \hspace{1mm}
	\subfloat[$\mathbf{\widehat{C}}_{32,\text{BAS}}$]{\includegraphics[scale = 0.625]{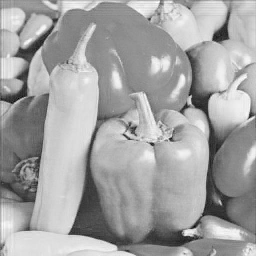}} \hspace{1mm}
	\subfloat[$\mathbf{\widehat{C}}_{32,\text{JAM}}$]{\includegraphics[scale = 0.625]{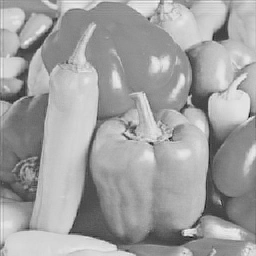}} \hspace{1mm}
	\subfloat[$\mathbf{\widehat{C}}_{32,\text{OCBSML}}$]{\includegraphics[scale = 0.625]{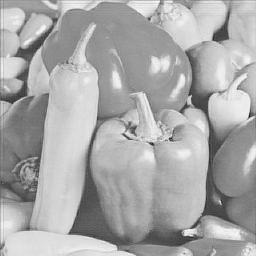}}\hspace{1mm}
	\subfloat[$\mathbf{\widehat{C}}_{\text{32},2}$]{\includegraphics[scale = 0.625]{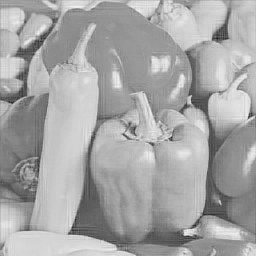}} \hspace{1mm}
	\subfloat[$\mathbf{\widehat{C}}_{16,5}^{(1)}$]{\includegraphics[scale = 0.625]{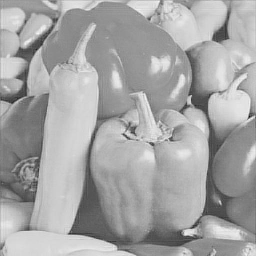}}
	\caption{Compressed \textit{Peppers} image for $N=32$ considering $r=205$ considering Method II.}
	\label{fig:pep-32_methodII}
\end{figure*}

\begin{figure*}[h!]
	\centering
	\subfloat[$\mathbf{C}_{64}$]{\includegraphics[scale = 0.625]{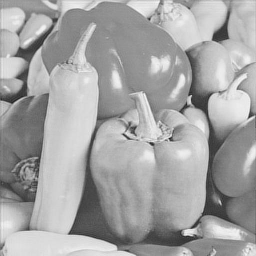}} \hspace{1mm}
	\subfloat[$\mathbf{\widehat{C}}_{64,\text{SDCT}}$]{\includegraphics[scale = 0.625]{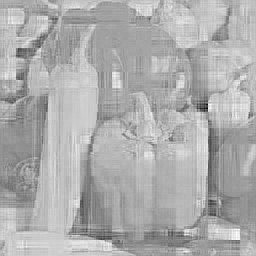}} \hspace{1mm}
	\subfloat[$\mathbf{\widehat{C}}_{64,\text{OCBSML}}$]{\includegraphics[scale = 0.625]{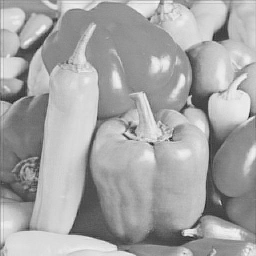}}\hspace{1mm}\\
	\subfloat[$\mathbf{\widehat{C}}_{\text{64},1}$]{\includegraphics[scale = 0.625]{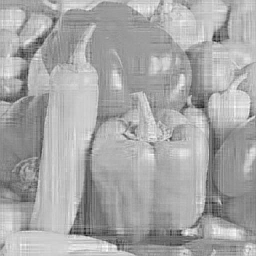}} \hspace{1mm}
	\subfloat[$\mathbf{\widehat{C}}_{16,5}^{(2)}$]{\includegraphics[scale = 0.625]{t641.eps}}\hspace{1mm}
	\subfloat[$\mathbf{\widehat{C}}_{32,2}^{(1)}$]{\includegraphics[scale = 0.625]{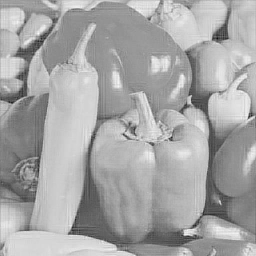}}
	\caption{Compressed \textit{Peppers} image for $N=64$ considering $r=820$ considering Method II.}
	\label{fig:pep-64_methodII}
\end{figure*}

\begin{table}[htb!]
	\centering
	\caption{Image quality measures of  compressed \textit{Peppers} image for $N=16$ $(r=50)$, $32$ $(r=205)$, and $64$ $(r=820)$.}
	\label{tab:image-measures}
	\begin{tabular}{clcccccccc}
		\toprule
		&&\multicolumn{3}{c}{Method I}&\multicolumn{3}{c}{Method II}\\\cmidrule(lr){3-5}\cmidrule{6-8}
		$N$
		& \multicolumn{1}{l}{\begin{tabular}[l]{@{}c@{}}Approximation \end{tabular}}
		& MSE & PSNR & MSSIM & MSE & PSNR & MSSIM
		\\
		\midrule
		16 &	$\mathbf{C}_{16}$&\textbf{32.3449}	&	\textbf{33.0328}	&	\textbf{0.9416}&\textbf{32.3449}	&	\textbf{33.0328}	&	\textbf{0.9416}	\\
		\hdashline
		&$\mathbf{\widehat{C}}_{\text{16},1}$&84.0657&28.8846&0.9107&79.6481&29.1190&0.9128	\\
		&$\mathbf{\widehat{C}}_{\text{16},2}$&76.4359&29.2978&0.9151&82.0878&28.9880&0.9111\\
		&$\mathbf{\widehat{C}}_{\text{16},3}$&37.9837&32.3348&0.9382&65.6444&29.9588&0.9244\\
		&$\mathbf{\widehat{C}}_{\text{16},4}$&38.0897&32.3227&0.9384&52.3516&30.9415&0.9312\\
		&$\mathbf{\widehat{C}}_{\text{16},5}$&38.6734&32.2567&0.9380&44.7897&31.6190&0.9349\\
		&$\mathbf{\widehat{C}}_{16,\text{SDCT}}$		&189.4979&25.3548&0.8690&161.0228&26.0619&0.8630\\

		&$\mathbf{\widehat{C}}_{16,\text{JAM}}$			& 59.6891	&	30.3718	&0.9242& 59.6891	&	30.3718	&0.9242	\\

		&$\mathbf{\widehat{C}}_{16,\text{BAS}}$		&	58.1761	&	30.4834	&	0.9248&	58.1761	&	30.4834	&	0.9248	\\

		&$\mathbf{\widehat{C}}_{16,\text{BCEM}}$		& 93.1786	&	28.4376	&0.9064&93.1786	&	28.4376	&0.9064	\\

		&$\mathbf{\widehat{C}}_{16,\text{SOBCM}}$		&115.2820	&	27.5132	&0.8897&115.2820	&	27.5132	&0.8897	\\

		&$\mathbf{\widehat{C}}_{16,\text{OCBSML}}$&49.3669&31.1964&0.9303&49.3669&31.1964&0.9303	\\

		\midrule
		32 &$\mathbf{C}_{32}$&\textbf{29.2562}	&\textbf{33.4686}&\textbf{0.9690}&\textbf{29.2562}	&\textbf{33.4686}&\textbf{0.9690}	\\

		\hdashline
				&$\mathbf{\widehat{C}}_{\text{32},1}$	&100.4120&28.1129&0.9394&201.1865&25.0948&0.9042	\\
				&$\mathbf{\widehat{C}}_{\text{32},2}$	&41.1957&31.9823&0.9643&72.8090&29.5090&0.9524\\
				&$\mathbf{\widehat{C}}_{16,5}^{(1)}$	&37.6285&32.3756&0.9653&43.8346&31.7126&0.9629\\
				&$\mathbf{\widehat{C}}_{32,\text{SDCT}}$&300.2389&23.3561&0.8782&410.2461&22.0004&0.8285	\\

		&$\mathbf{\widehat{C}}_{32,\text{JAM}}$	&58.4496&30.4630&0.9554&58.4496&30.4630&0.9554	\\

		&$\mathbf{\widehat{C}}_{32,\text{BAS}}$	&61.6647&30.2304&0.9524&61.6647&30.2304&0.9524	\\

		&$\mathbf{\widehat{C}}_{32,\text{OCBSML}}$&47.9872&31.3196&0.9591&47.9872&31.3196&0.9591	\\

		\midrule
		64	&$\mathbf{C}_{64}$	& \textbf{28.1330}&	\textbf{33.6386}&	\textbf{0.9880}& \textbf{28.1330}&	\textbf{33.6386}&	\textbf{0.9880}	\\

		\hdashline
				&$\mathbf{\widehat{C}}_{\text{64},1}$	&177.9596&25.6276&0.9513&282.3732&23.6226&0.9231\\
				&$\mathbf{\widehat{C}}_{16,5}^{(2)}$	&37.1861&32.4270&0.9851&43.4505&31.7509&0.9834\\
				&$\mathbf{\widehat{C}}_{32,2}^{(1)}$	&40.5326&32.0528&0.9842&73.1704&29.4874&0.9751\\
				&$\mathbf{\widehat{C}}_{64,\text{SDCT}}$&510.5425&21.0505&0.8846&617.3670&20.2254&0.8380\\
				&$\mathbf{\widehat{C}}_{64,\text{OCBSML}}$&47.5035&31.3635&0.9817&47.5035&31.3635&0.9817\\
		\bottomrule
	\end{tabular}
\end{table}

\subsection{Quantitative analysis}
In this analysis, we considered the measurements of the selected $45$ $8$-bit
images obtained from a public image bank~\cite{uscsipi}.
Each image was compressed considering the initial $r$ coefficient~(matrix elements ordered according to the zig-zag pattern), $r \in \left\{0,1,\ldots N^2\right\}$,
and assessed by the quality measures.

\begin{figure*}[h!]
\centering
\psfrag{r}[l][l][0.5]{$r$}
\psfrag{q}[l][l][0.5]{$\mathbf{C}_{16}$}
\psfrag{V2}[l][l][0.5]{$\mathbf{\widehat{C}}_{16,\text{BAS}}$}
\psfrag{V3}[l][l][0.5]{$\mathbf{\widehat{C}}_{16,\text{JAM}}$}
\psfrag{V4}[l][l][0.5]{$\mathbf{\widehat{C}}_{16,\text{OCBSML}}$}
\psfrag{V5}[l][l][0.5]{$\mathbf{\widehat{C}}_{\text{16},5}$}
\subfloat{\includegraphics[scale = 0.4]{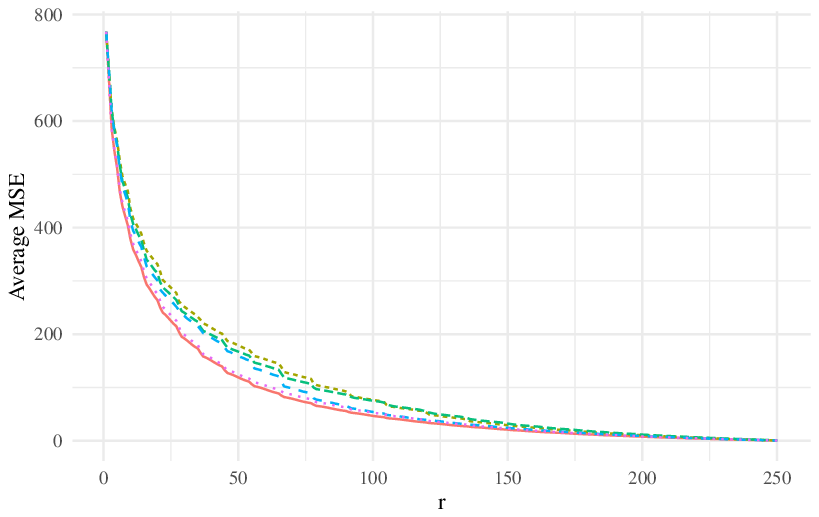}}
\subfloat{\includegraphics[scale = 0.4]{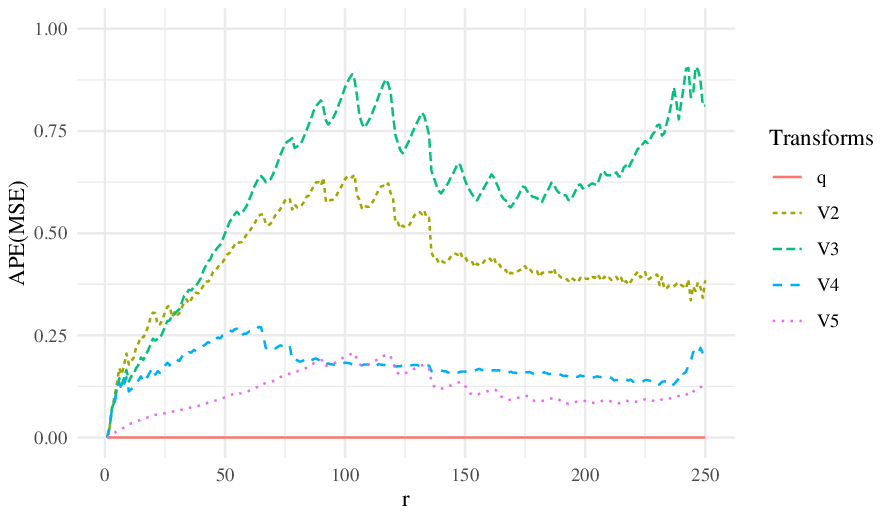}}\\
\subfloat{\includegraphics[scale = 0.4]{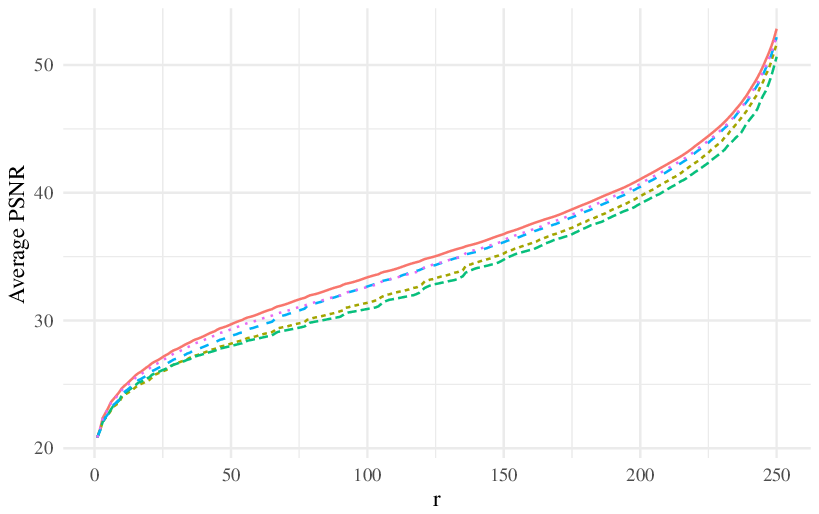}}
\subfloat{\includegraphics[scale = 0.4]{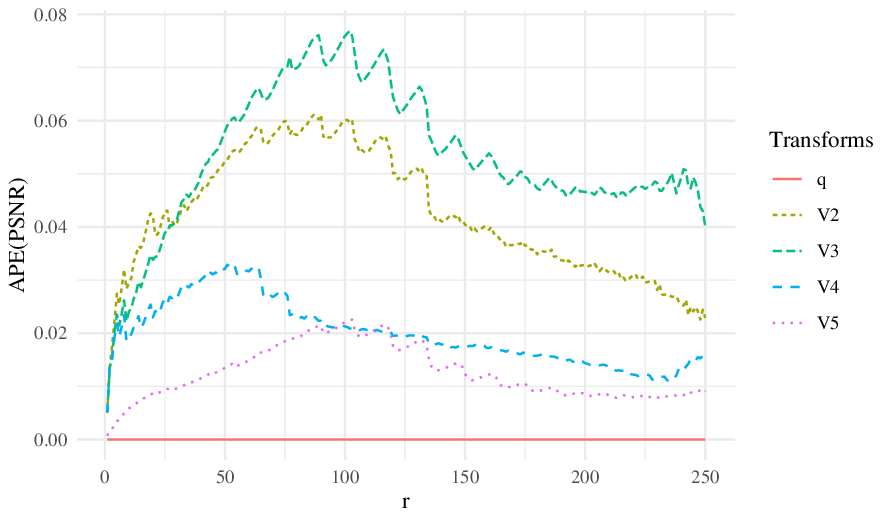}}\\
\subfloat{\includegraphics[scale = 0.4]{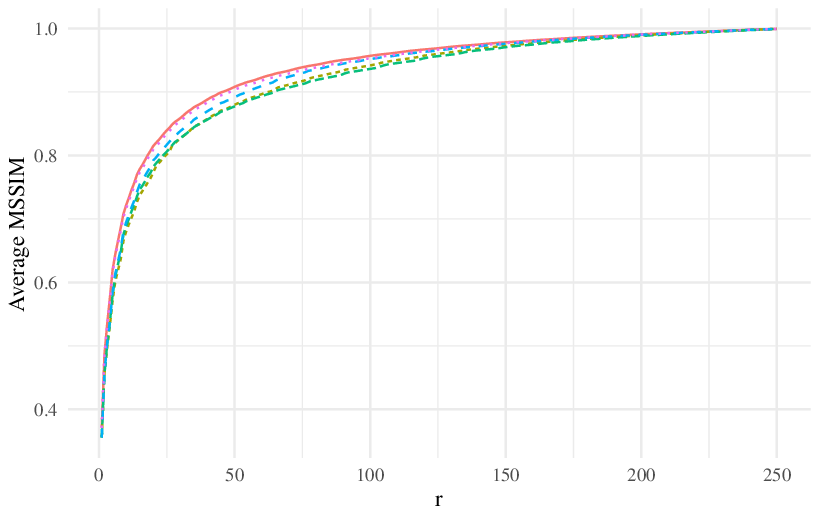}}
\subfloat{\includegraphics[scale = 0.4]{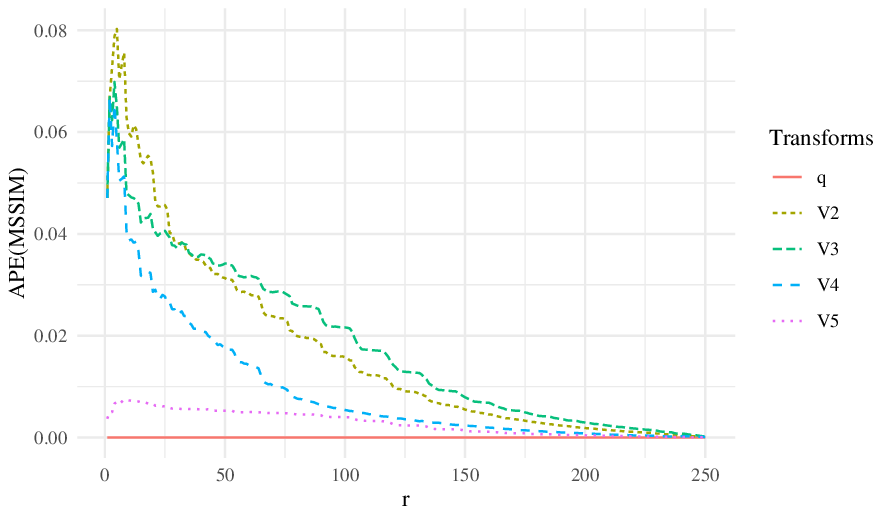}}
\caption{Average curves for the MSE, PSNR, and  MSSIM
	for the $16$-point DCT and approximations based on Method I.}
\label{fig:avg_curves_DCT_16_methodI}
\end{figure*}

\begin{figure*}[h!]
	\centering
	\psfrag{r}[l][l][0.5]{$r$}
	\psfrag{q}[l][l][0.5]{$\mathbf{C}_{16}$}
	\psfrag{V2}[l][l][0.5]{$\mathbf{\widehat{C}}_{16,\text{BAS}}$}
	\psfrag{V3}[l][l][0.5]{$\mathbf{\widehat{C}}_{16,\text{JAM}}$}
	\psfrag{V4}[l][l][0.5]{$\mathbf{\widehat{C}}_{16,\text{OCBSML}}$}
	\psfrag{V5}[l][l][0.5]{$\mathbf{\widehat{C}}_{\text{16},5}$}
	\subfloat{\includegraphics[scale = 0.4]{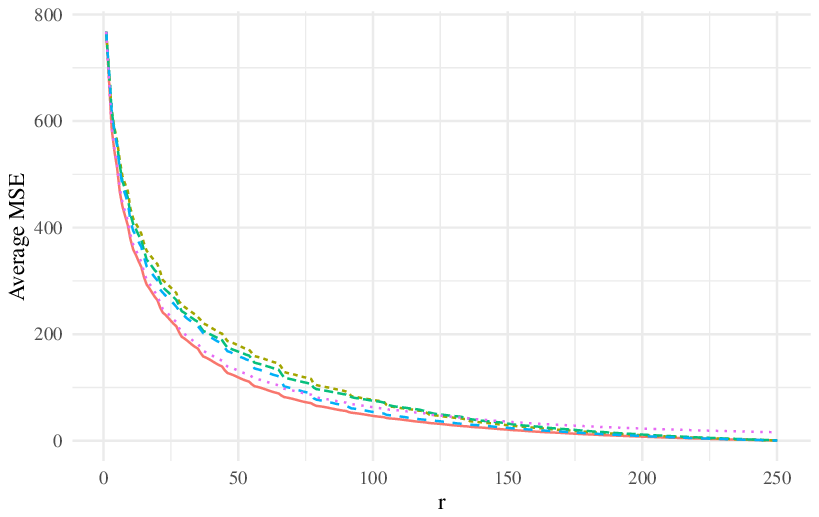}}
	\subfloat{\includegraphics[scale = 0.4]{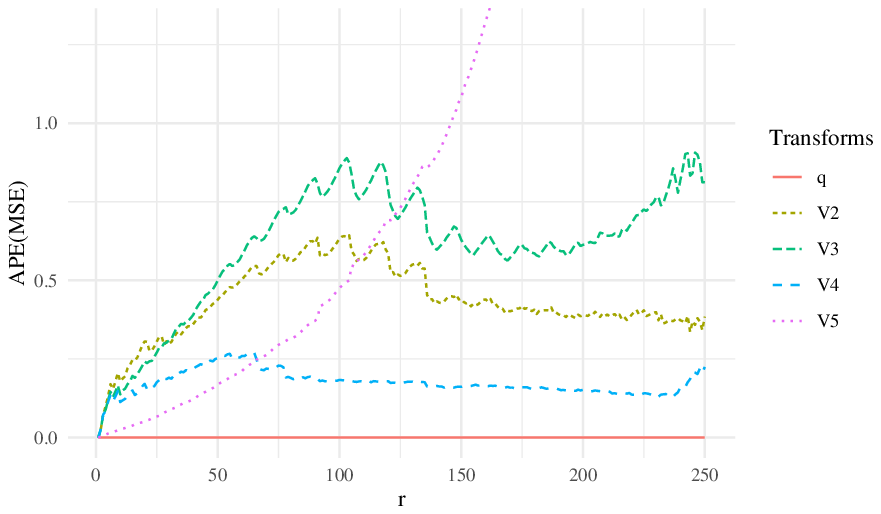}}\\
	\subfloat{\includegraphics[scale = 0.4]{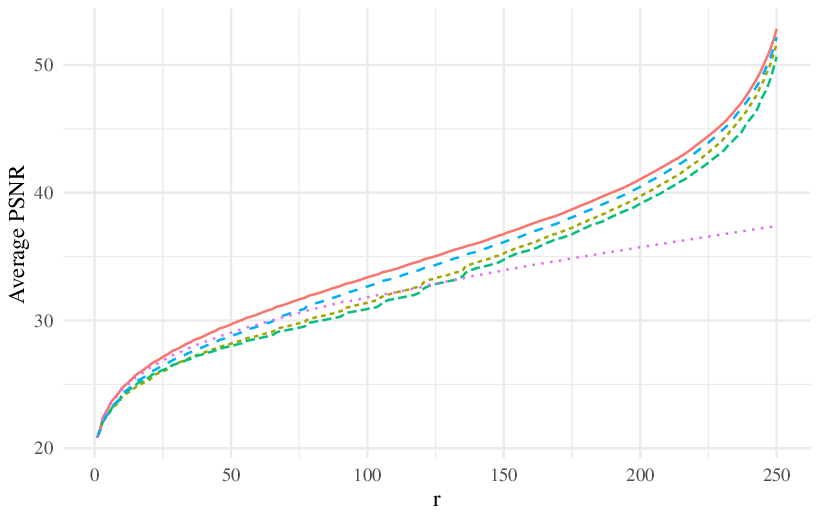}}
	\subfloat{\includegraphics[scale = 0.4]{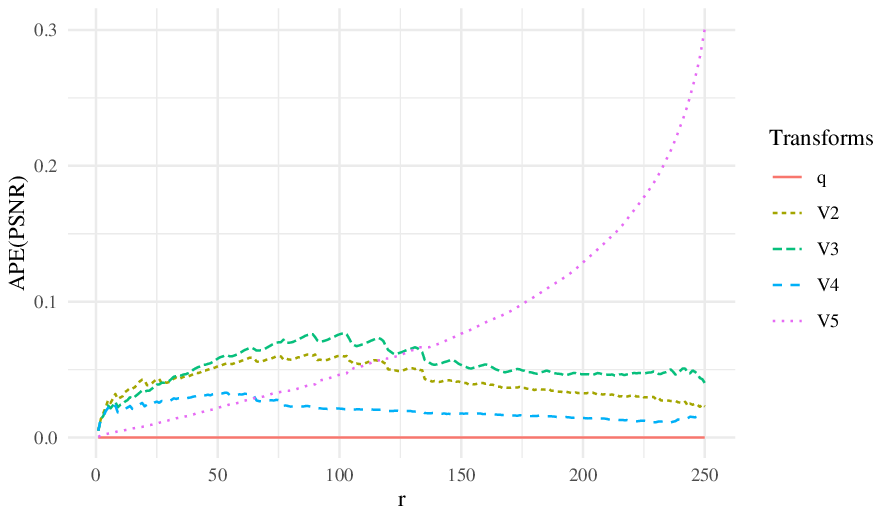}}\\
	\subfloat{\includegraphics[scale = 0.4]{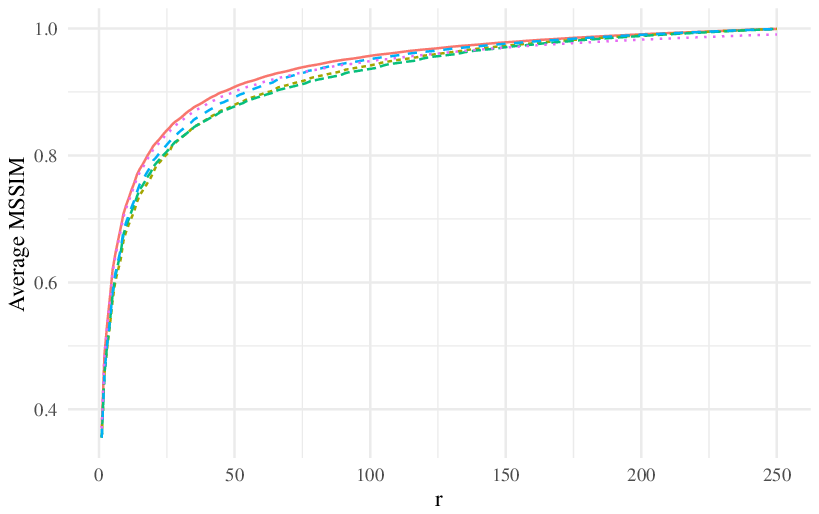}}
	\subfloat{\includegraphics[scale = 0.4]{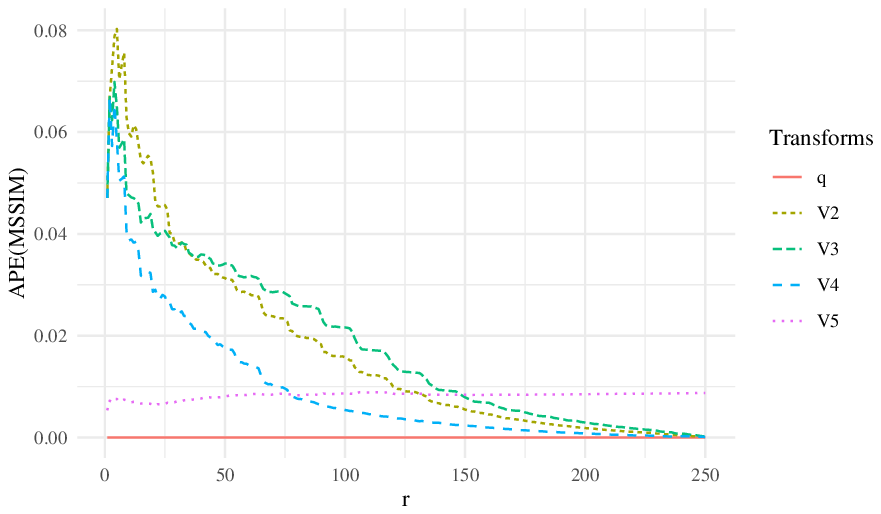}}
	\caption{Average curves for the MSE, PSNR, and  MSSIM
		for the $16$-point DCT and approximations based on Method II.}
	\label{fig:avg_curves_DCT_16_methodII}
\end{figure*}

\begin{figure*}[h!]
\centering
\psfrag{r}[l][l][0.5]{$r$}
\psfrag{q}[l][l][0.5]{$\mathbf{C}_{32}$}
\psfrag{V2}[l][l][0.5]{$\mathbf{\widehat{C}}_{32,\text{BAS}}$}
\psfrag{V3}[l][l][0.5]{$\mathbf{\widehat{C}}_{32,\text{JAM}}$}
\psfrag{V4}[l][l][0.5]{$\mathbf{\widehat{C}}_{32,\text{OCBSML}}$}
\psfrag{V5}[l][l][0.5]{$\mathbf{\widehat{C}}_{\text{32},2}$}
\psfrag{V6}[l][l][0.5]{$\mathbf{\widehat{C}}_{16,5}^{(1)}$}
\subfloat{\includegraphics[scale = 0.4]{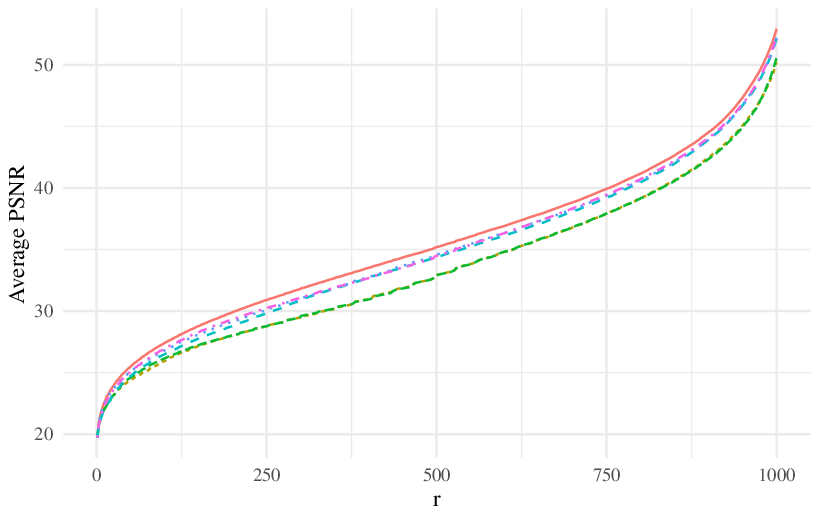}}
\subfloat{\includegraphics[scale = 0.4]{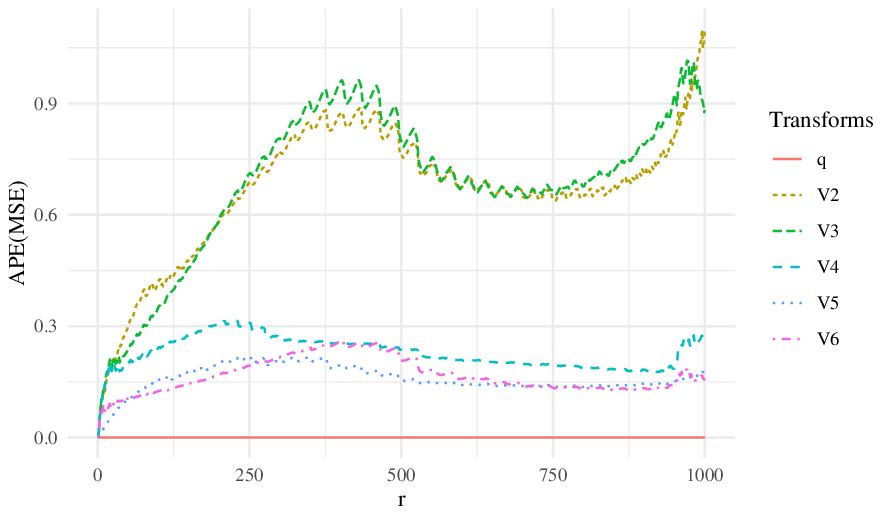}}\\
\subfloat{\includegraphics[scale = 0.4]{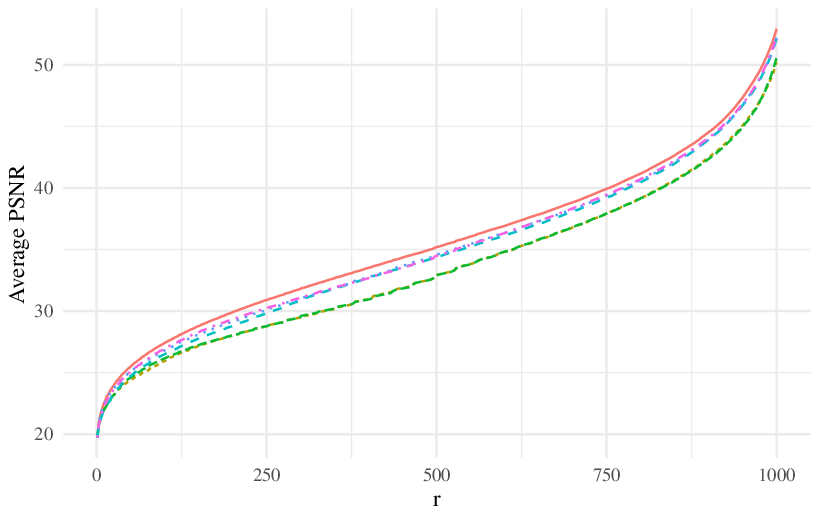}}
\subfloat{\includegraphics[scale = 0.4]{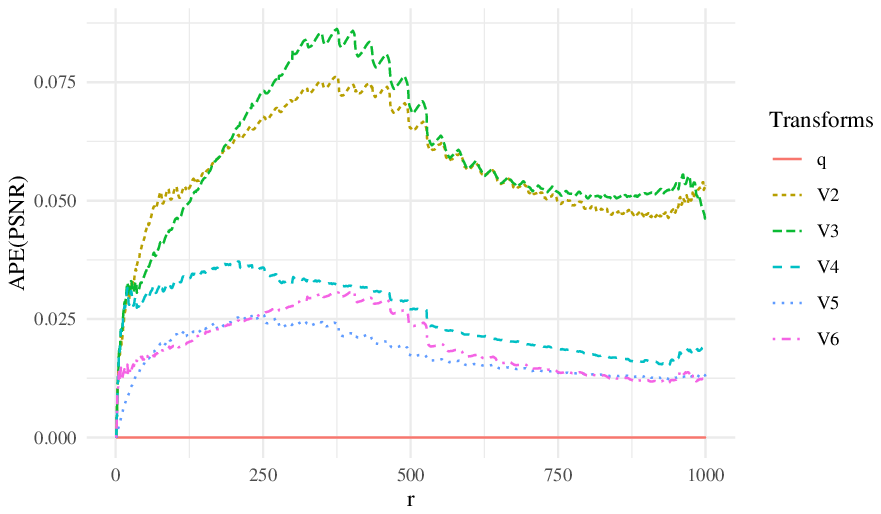}}\\
\subfloat{\includegraphics[scale = 0.4]{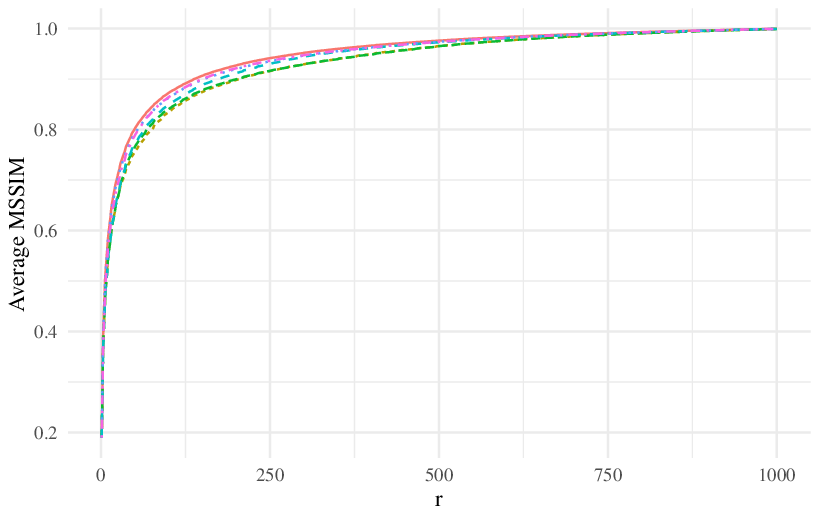}}
\subfloat{\includegraphics[scale = 0.4]{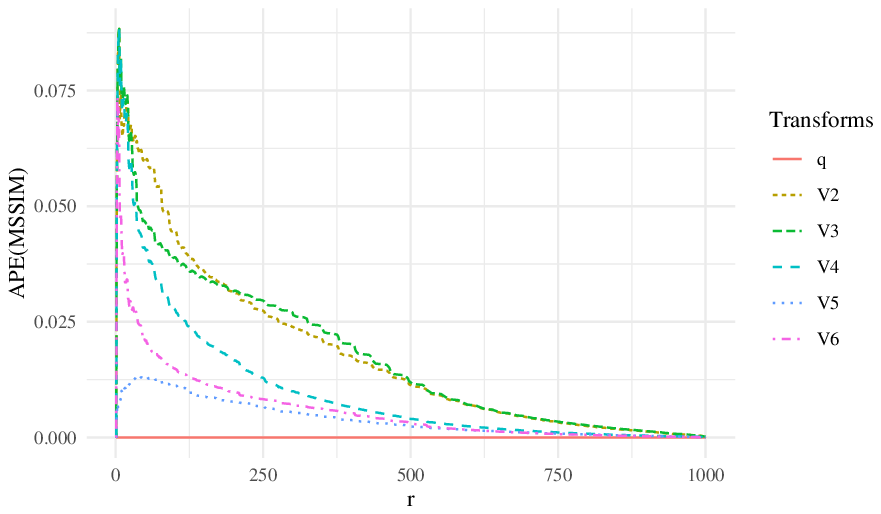}}
\caption{Average curves for the MSE, PSNR, and MSSIM
	for the $32$-point DCT and approximations based on Method I.}
\label{fig:avg_curves_DCT_32_methodI}
\end{figure*}

\begin{figure*}[h!]
	\centering
	\psfrag{r}[l][l][0.5]{$r$}
	\psfrag{q}[l][l][0.5]{$\mathbf{C}_{32}$}
	\psfrag{V2}[l][l][0.5]{$\mathbf{\widehat{C}}_{32,\text{BAS}}$}
	\psfrag{V3}[l][l][0.5]{$\mathbf{\widehat{C}}_{32,\text{JAM}}$}
	\psfrag{V4}[l][l][0.5]{$\mathbf{\widehat{C}}_{32,\text{OCBSML}}$}
	\psfrag{V5}[l][l][0.5]{$\mathbf{\widehat{C}}_{\text{32},2}$}
	\psfrag{V6}[l][l][0.5]{$\mathbf{\widehat{C}}_{16,5}^{(1)}$}
	\subfloat{\includegraphics[scale = 0.4]{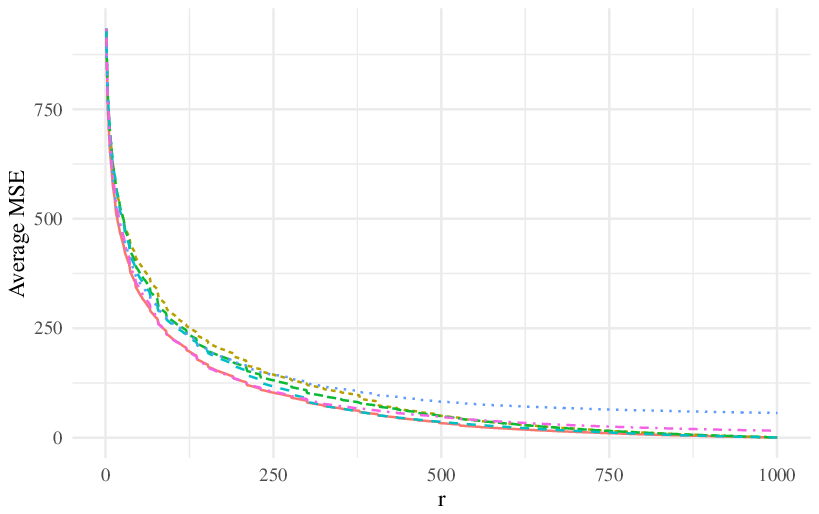}}
	\subfloat{\includegraphics[scale = 0.4]{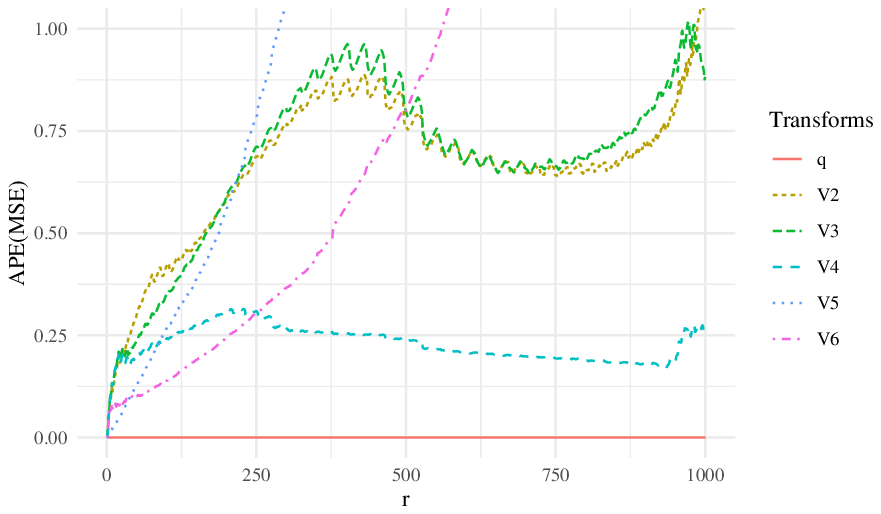}}\\
	\subfloat{\includegraphics[scale = 0.4]{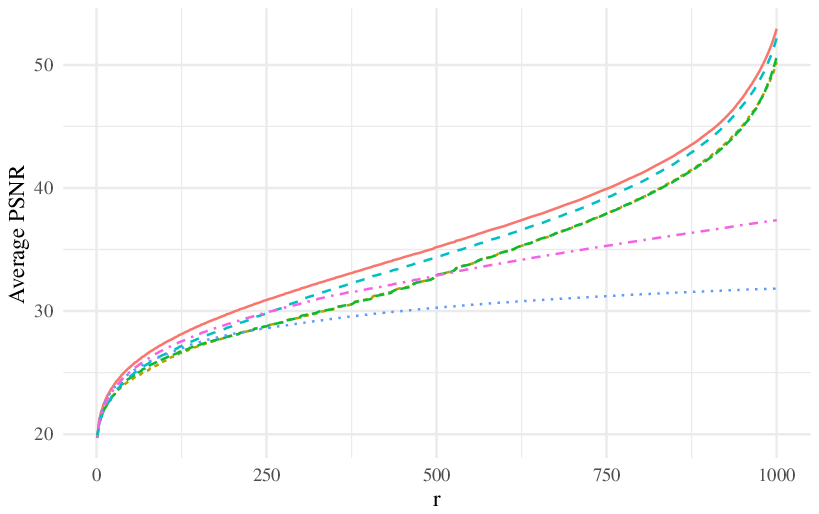}}
	\subfloat{\includegraphics[scale = 0.4]{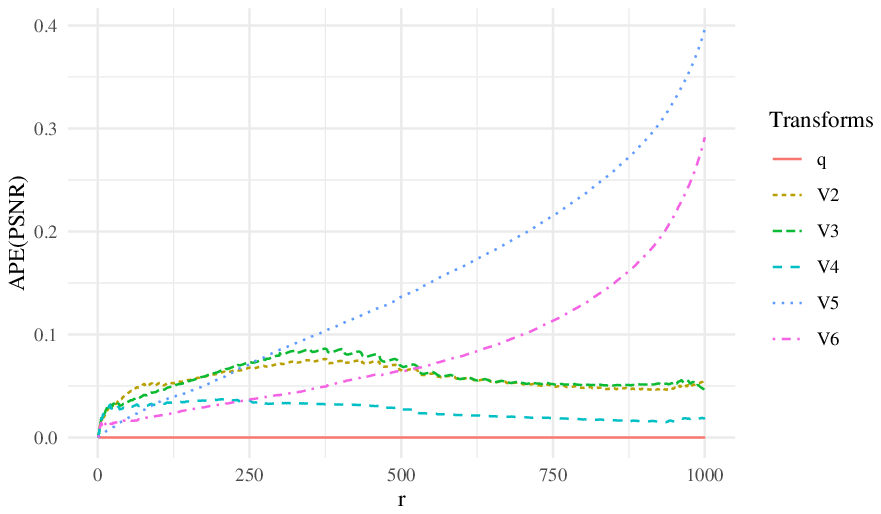}}\\
	\subfloat{\includegraphics[scale = 0.4]{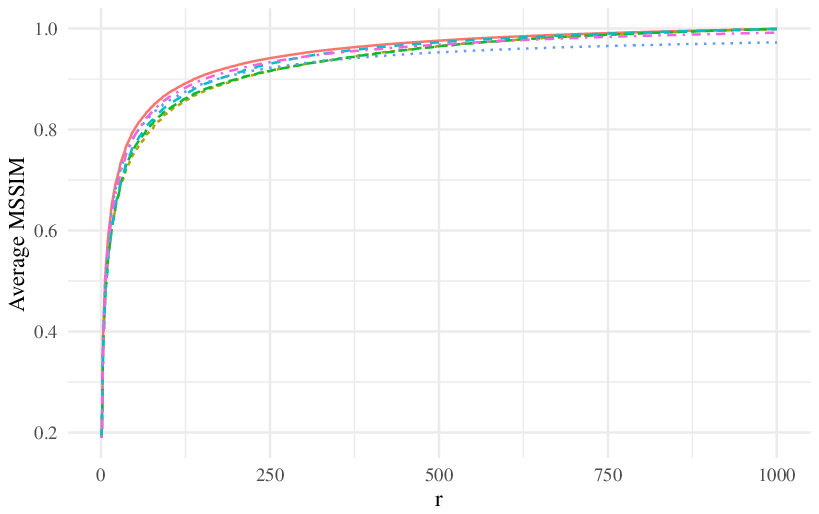}}
	\subfloat{\includegraphics[scale = 0.4]{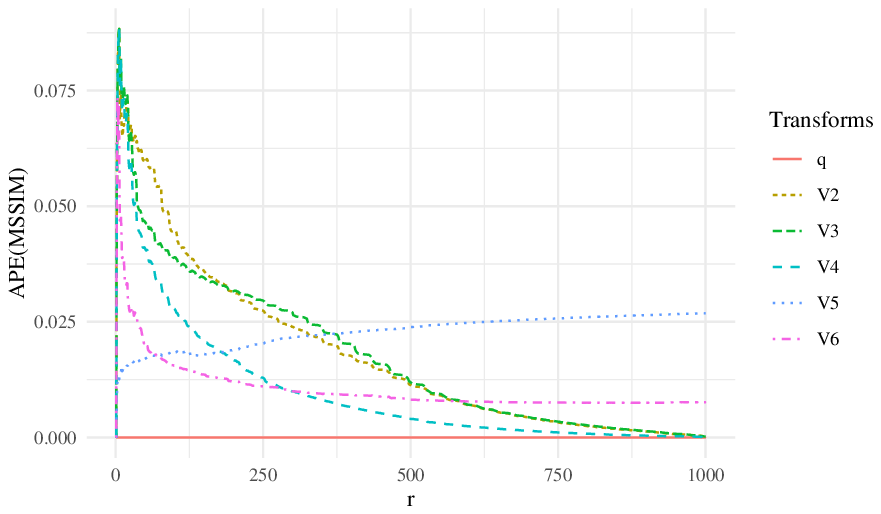}}
	\caption{Average curves for the MSE, PSNR, and MSSIM
		for the $32$-point DCT and approximations based on Method II.}
	\label{fig:avg_curves_DCT_32_methodII}
\end{figure*}

\begin{figure*}[h!]
\centering
\psfrag{r}[l][l][0.5]{$r$}
\psfrag{q}[l][l][0.5]{$\mathbf{C}_{64}$}
\psfrag{V2}[l][l][0.5]{$\mathbf{\widehat{C}}_{64,\text{OCBSML}}$}
\psfrag{V3}[l][l][0.5]{$\mathbf{\widehat{C}}_{\text{64},1}$}
\psfrag{V4}[l][l][0.5]{$\mathbf{\widehat{C}}_{16,5}^{(2)}$}
\psfrag{V5}[l][l][0.5]{$\mathbf{\widehat{C}}_{32,2}^{(1)}$}
\subfloat{\includegraphics[scale = 0.4]{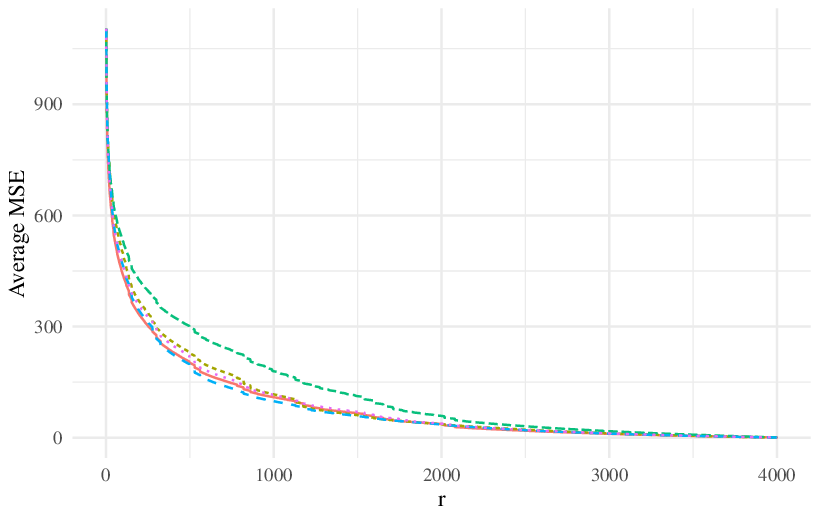}}
\subfloat{\includegraphics[scale = 0.4]{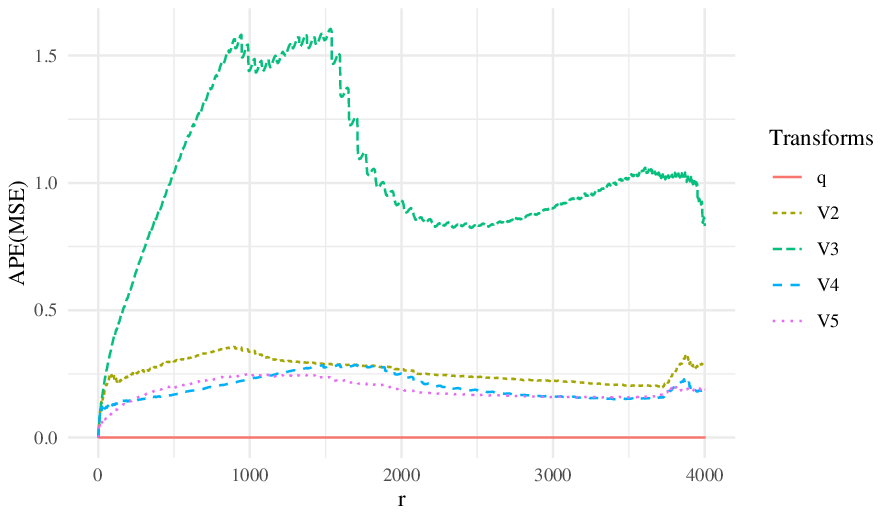}}\\
\subfloat{\includegraphics[scale = 0.4]{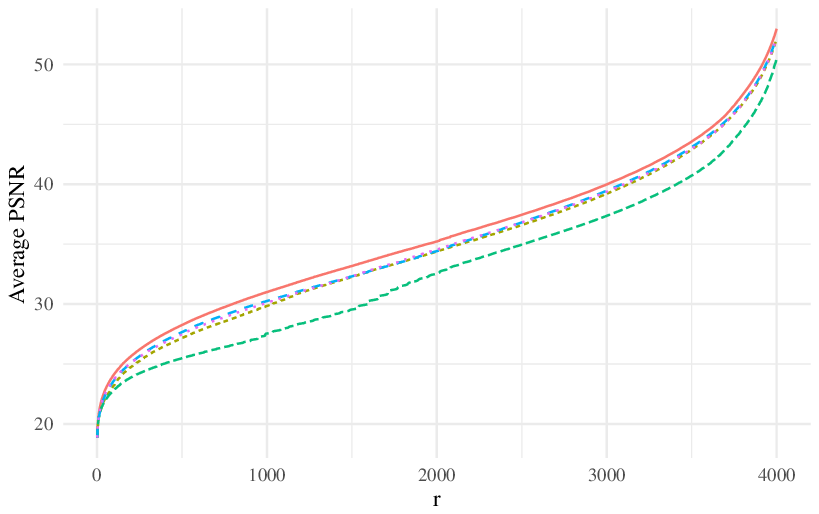}}
\subfloat{\includegraphics[scale = 0.4]{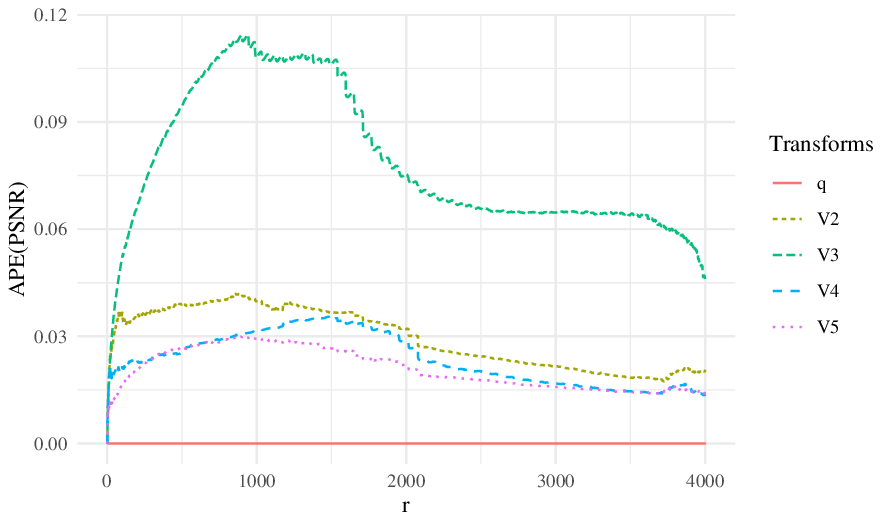}} \\
\subfloat{\includegraphics[scale = 0.4]{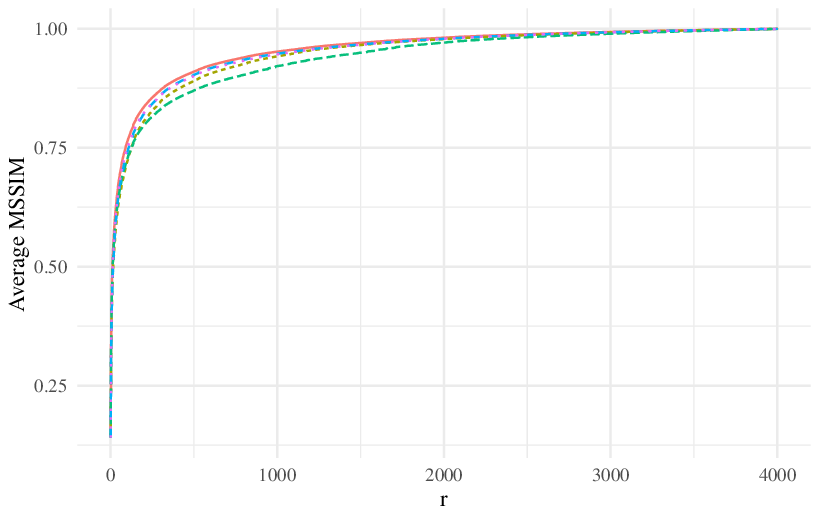}}
\subfloat{\includegraphics[scale = 0.4]{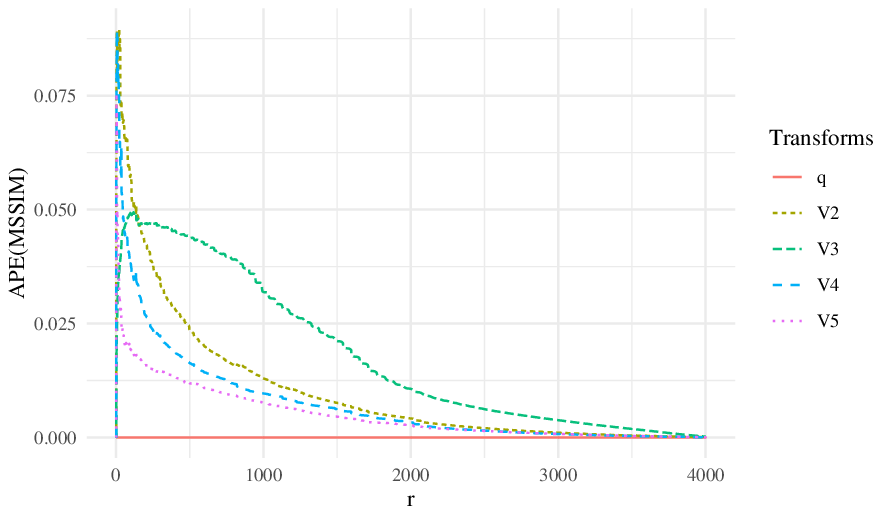}}
\caption{Average curves for the MSE, PSNR, and MSSIM
	for the $64$-point DCT and approximations based on Method I.}
\label{fig:avg_curves_DCT_64_methodI}
\end{figure*}

\begin{figure*}[h!]
	\centering
	\psfrag{r}[l][l][0.5]{$r$}
	\psfrag{q}[l][l][0.5]{$\mathbf{C}_{64}$}
	\psfrag{V2}[l][l][0.5]{$\mathbf{\widehat{C}}_{64,\text{OCBSML}}$}
	\psfrag{V3}[l][l][0.5]{$\mathbf{\widehat{C}}_{\text{64},1}$}
	\psfrag{V4}[l][l][0.5]{$\mathbf{\widehat{C}}_{16,5}^{(2)}$}
	\psfrag{V5}[l][l][0.5]{$\mathbf{\widehat{C}}_{32,2}^{(1)}$}
	\subfloat{\includegraphics[scale = 0.4]{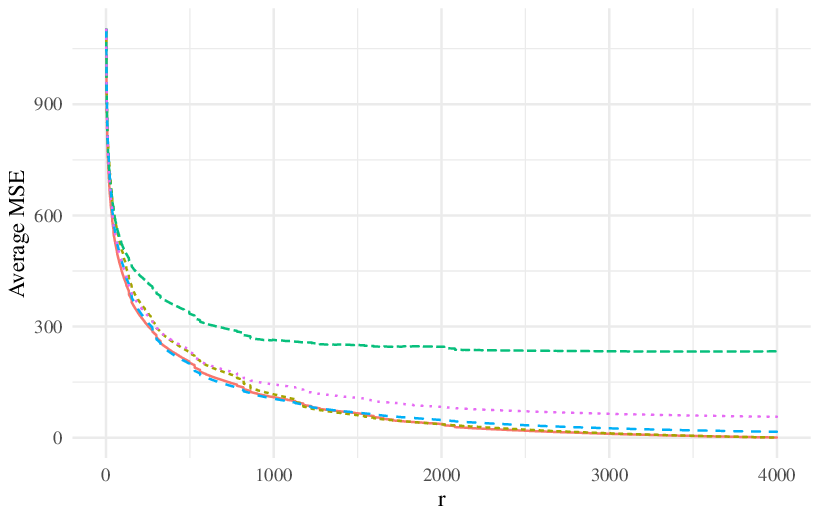}}
	\subfloat{\includegraphics[scale = 0.4]{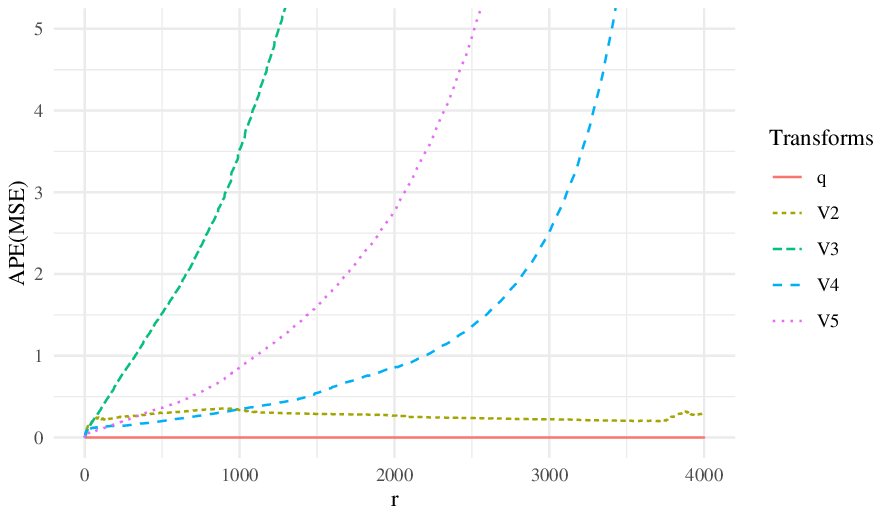}}\\
	\subfloat{\includegraphics[scale = 0.4]{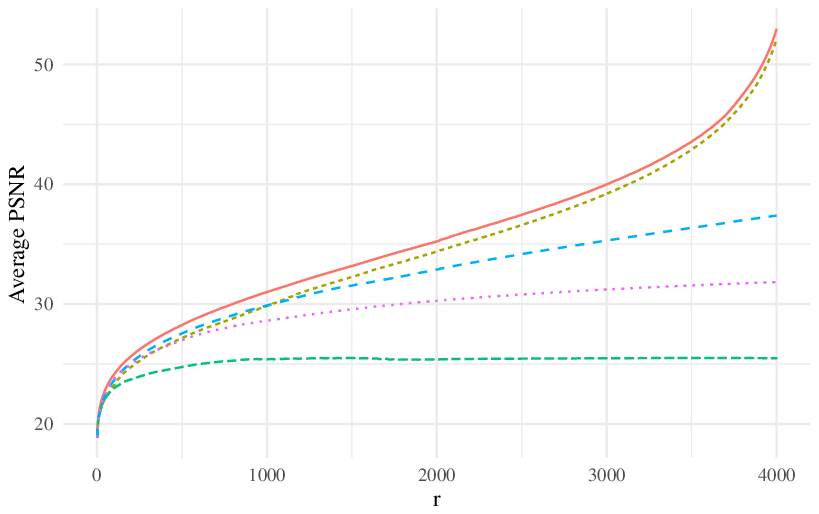}}
	\subfloat{\includegraphics[scale = 0.4]{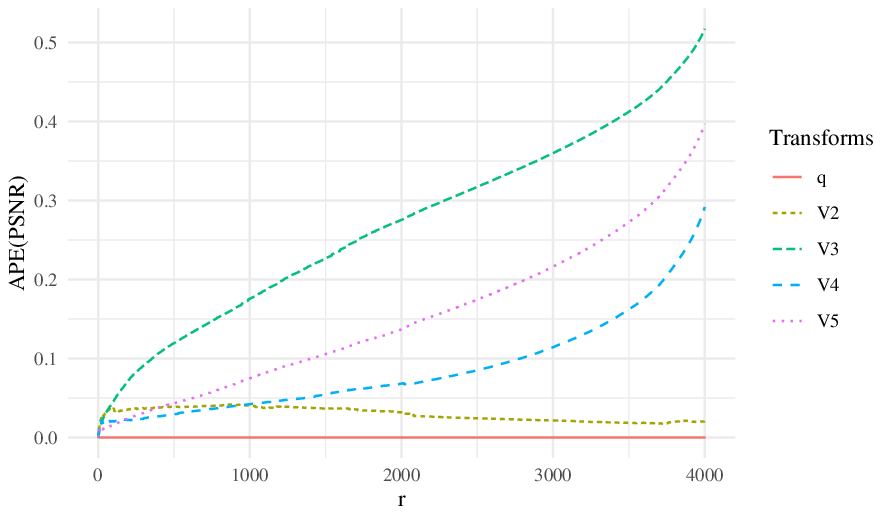}} \\
	\subfloat{\includegraphics[scale = 0.4]{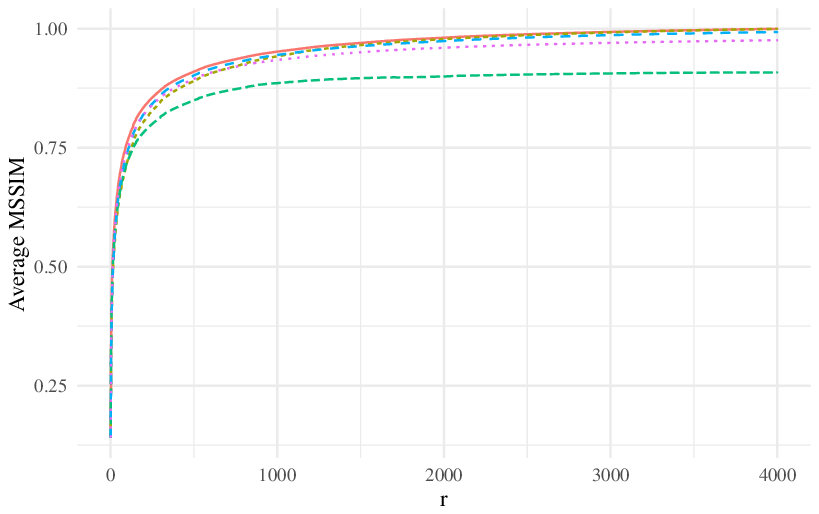}}
	\subfloat{\includegraphics[scale = 0.4]{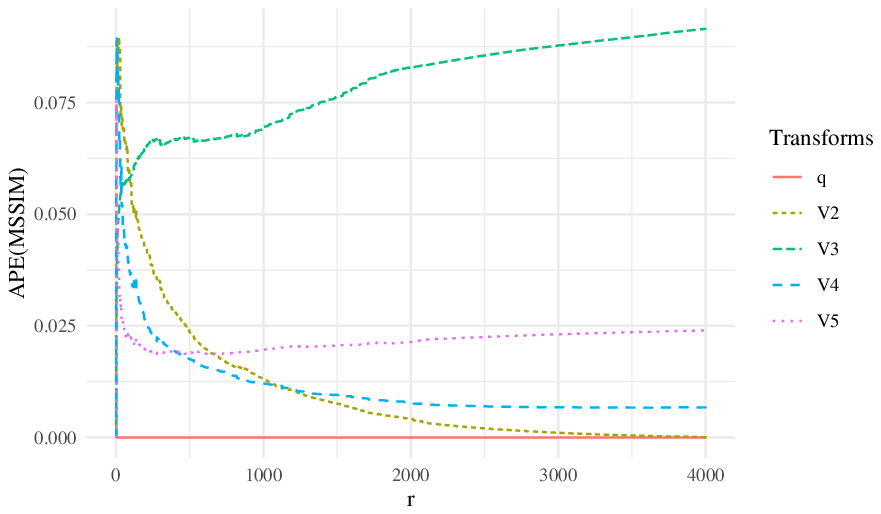}}
	\caption{Average curves for the MSE, PSNR, and MSSIM
		for the $64$-point DCT and approximations based on Method II.}
	\label{fig:avg_curves_DCT_64_methodII}
\end{figure*}

Fig.~\ref{fig:avg_curves_DCT_16_methodI},
\ref{fig:avg_curves_DCT_16_methodII},
\ref{fig:avg_curves_DCT_32_methodI},
\ref{fig:avg_curves_DCT_32_methodII},
\ref{fig:avg_curves_DCT_64_methodI},
and~\ref{fig:avg_curves_DCT_64_methodII}
show the average curves from the $45$ images for the MSE, PSNR, and MSSIM based on Method I and II.
For better visualization, we adopted the absolute percentage error (APE) relative to the DCT:
\begin{eqnarray*}
	\operatorname{APE}(\mu) = \left| \frac{\mu(\mathbf{C}_N) - \mu(
		\mathbf{\widehat{C}}_N)}{\mu(\mathbf{C}_N)} \right|,
	\quad \mu \in \{\text{MSE}, \text{PSNR}, \text{MSSIM}\},
\end{eqnarray*}
where $\mu(\mathbf{C}_N)$ and $\mu(\mathbf{\widehat{C}}_N)$ indicate the
measurements according to the exact and approximate $N$-point DCT, respectively.
For this evaluation, we considered the optimal proposed transforms  $\mathbf{\widehat{C}}_{\text{16},5}$, $\mathbf{\widehat{C}}_{\text{32},2}$, $\mathbf{\widehat{C}}_{16,5}^{(1)}$, $\mathbf{\widehat{C}}_{\text{64},1}$, $\mathbf{\widehat{C}}_{16,5}^{(2)}$, and $\mathbf{\widehat{C}}_{32,2}^{(1)}$ and compared with the exact DCT ($\mathbf{C}_N$) and the three best approximations according to Table~\ref{tab:apps_measures}: $\mathbf{\widehat{C}}_{N,\text{BAS}}$, $\mathbf{\widehat{C}}_{N,\text{JAM}}$, and $\mathbf{\widehat{C}}_{N,\text{OCBSML}}$.
\textbf{The proposed approximations are the ones with the best results when contrasted with competing DCT approximations
	for $N = 16$, $32$, and $64$.}

\FloatBarrier
\section{Conclusions}\label{S:conclusion}
This paper introduces low-complexity approximations for the DCT of lengths 16, 32, and 64, based on the method proposed in~\cite{oliveira2019low}, which approximates
the rows of the exact and  the approximate matrix transform according the angle between them.
The proposed transforms were assessed according to classical figures of merit and showed an outstanding performance when compared to the DCT approximations already archived in the literature. Fast algorithms were also derived for the best transforms of each blocklength, which further reduced their arithmetic costs. Only additions and bit-shifting operations are necessary  for their computation.
Besides, the transforms were evaluated in the context of image compression. \textit{The experiments demonstrated that the proposed transforms outperformed all the considered DCT approximations already known in the literature for $N = 16$, $32$, and $64$.}

\section*{Acknowledgments}
We gratefully acknowledge partial financial support from  \textit{Coordena\c c\~ao de Aperfei\c coamento de Pessoal de N\'ivel Superior (CAPES)}, \textit{Conselho Nacional de Desenvolvimento Cient\'ifico e Tecnol\'ogico (CNPq)}, and \textit{Funda\c c\~ao de Amparo a Ci\^encia e Tecnologia de Pernambuco (FACEPE)}, Brazil.

\appendix
\section{Matrix Factorization}
The matrices used in the Section \ref{S:FastAlgo} are detailed here. First, consider the following butterfly-structure:
\begin{equation*}
	\mathbf{B}_{N}=
		\begin{bmatrix}
			{\mathbf{I}}_{\frac{N}{2}}&\bar{\mathbf{I}}_{\frac{N}{2}}\\
			-\bar{\mathbf{I}}_{\frac{N}{2}}&{\mathbf{I}}_{\frac{N}{2}}
		\end{bmatrix}.
\end{equation*}

For the sake of brevity, cycle notation ~\cite{dummit2004abstract,zassenhaus2013theory,rotman2012introduction} is used
to present the permutation matrices.
The resulting permutation matrix is obtained by permuting the columns of the identity matrix following the cycle mapping zero indexed.

\subsection{Matrix Factorization for $N=16$}
The low-complexity matrix $\mathbf{T}_{16,5}$ can be represented as:
\begin{equation*}
	\mathbf{T}_{16,5}=\mathbf{P}_{16}\cdot\mathbf{M}_{16}
	\cdot
	\left[
	\begin{array}{cc}
		\mathbf{B}_{2}&\\
		&{\mathbf{I}}_{14}
	\end{array}
	\right]
	\cdot
	\left[
	\begin{array}{cc}
		\mathbf{B}_{4}&\\
		&{\mathbf{I}}_{12}
	\end{array}
	\right]
	\cdot
	\left[
	\begin{array}{cc}
		\mathbf{B}_{8}&\\
		&{\mathbf{I}}_{8}
	\end{array}\right]\cdot\mathbf{B}_{16}
\end{equation*}
where
\begin{equation*}
	{\mathbf{P}}_{16}=\scalebox{0.9}{(1\;8)(2\;4)(3\;12\;9)(5\;6\;10)(7\;14\;13\;11)},
\end{equation*}

\begin{equation*}
	\footnotesize
	{\mathbf{M}}_{16}=
	\left[\hspace{-0.1cm}\begin{rsmallmatrix}\\
		1 &  &  &  &  &  &  &  &  &  &  &  &  &  &  &  \\
		& -1 &  &  &  &  &  &  &  &  &  &  &  &  &  &  \\
		&  & -1 & -2 &  &  &  &  &  &  &  &  &  &  &  &  \\
		&  & 2 & -1 &  &  &  &  &  &  &  &  &  &  &  &  \\
		&  &  &  & -\frac{1}{2} & -1 & -2 & -2 &  &  &  &  &  &  &  &  \\
		&  &  &  & 1 & 2 & \frac{1}{2} & -2 &  &  &  &  &  &  &  &  \\
		&  &  &  & -2 & -\frac{1}{2} & 2 & -1 &  &  &  &  &  &  &  &  \\
		&  &  &  & 2 & -2 & 1 & -\frac{1}{2} &  &  &  &  &  &  &  &  \\
		&  &  &  &  &  &  &  & -\frac{1}{4} & -\frac{1}{2} & -1 & -1 & -2 & -2 & -2 & -2 \\
		&  &  &  &  &  &  &  & \frac{1}{2} & 2 & 2 & 2 & 1 & -\frac{1}{4} & -1 & -2 \\
		&  &  &  &  &  &  &  & -1 & -2 & -1 & \frac{1}{2} & 2 & 2 & -\frac{1}{4} & -2 \\
		&  &  &  &  &  &  &  & 1 & 2 & -\frac{1}{2} & -2 & -\frac{1}{4} & 2 & 1 & -2 \\
		&  &  &  &  &  &  &  & -2 & -1 & 2 & \frac{1}{4} & -2 & \frac{1}{2} & 2 & -1 \\
		&  &  &  &  &  &  &  & 2 & -\frac{1}{4} & -2 & 2 & -\frac{1}{2} & -1 & 2 & -1 \\
		&  &  &  &  &  &  &  & -2 & 1 & -\frac{1}{4} & -1 & 2 & -2 & 2 & -\frac{1}{2} \\
		&  &  &  &  &  &  &  & 2 & -2 & 2 & -2 & 1 & -1 & \frac{1}{2} & -\frac{1}{4} \\
	\end{rsmallmatrix}\hspace{-0.1cm}\right].
\end{equation*}

\subsection{Matrix Factorization for $N=32$}
The low-complexity matrix $\mathbf{T}_{32,2}$ can be represented as:
\begin{eqnarray*}
	\mathbf{T}_{32,2}=\mathbf{P}_{32}\cdot
	\left[
	\begin{array}{cc}
		\mathbf{L}_{1}&\\
		&{\mathbf{L}}_{2}
	\end{array}
	\right]
	\cdot
	\left[
	\begin{array}{cc}
		\mathbf{B}_{2}&\\
		&{\mathbf{I}}_{30}
	\end{array}
	\right]
	\cdot
	\left[
	\begin{array}{cc}
		\mathbf{B}_{4}&\\
		&{\mathbf{I}}_{28}
	\end{array}
	\right]
	\cdot
	\left[
	\begin{array}{cc}
		\mathbf{B}_{8}&\\
		&{\mathbf{I}}_{24}
	\end{array}
	\right] \\
	\cdot
	\left[
	\begin{array}{cc}
		\mathbf{B}_{16}&\\
		&{\mathbf{I}}_{16}
	\end{array}
	\right]
	\cdot
	\mathbf{B}_{32}
\end{eqnarray*}
where
\begin{equation*}
	\small
	{\mathbf{P}}_{32}=\scalebox{0.9}{(1\;16\;5\;20\;13\;26\;25\;23\;19\;11\;18\;9\;10\;14\;30)(2\;8\;6\;28\;29\;31\;3\;24\;21\;15)(4\;12\;22\;17\;7)},
\end{equation*}

\begin{equation*}
	\footnotesize
	{\mathbf{L}}_1=
	\left[\hspace{-0.1cm}\begin{rsmallmatrix}\\
		1 &  &  &  &  &  &  &  &  &  &  &  &  &  &  &  \\
		& -1 &  &  &  &  &  &  &  &  &  &  &  &  &  &  \\
		&  & -\frac{1}{2} & -1 &  &  &  &  &  &  &  &  &  &  &  &  &    \\
		&  & 1 & -\frac{1}{2} &  &  &  &  &  &  &  &  &  &  &  &  &  \\
		&  &  &  & \frac{1}{2} & 1 &  & -1 &  &  &  &  &  &  &  &  &  \\
		&  &  &  & -1 &  & 1 & -\frac{1}{2} &  &  &  &  &  &  &  &  &   \\
		&  &  &  & 1 & -1 & \frac{1}{2} &  &  &  &  &  &  &  &  &  &     \\
		&  &  &  &  & -\frac{1}{2} & -1 & -1 &  &  &  &  &  &  &  &  &                               \\
		&  &  &  &  &  &  &  & \frac{1}{2} & 1 & 1 & 1 & \frac{1}{2} &  & -\frac{1}{2} & -1 &  \\
		&  &  &  &  &  &  &  & -\frac{1}{2} & -1 & -\frac{1}{2} & \frac{1}{2} & 1 & 1 &  & -1 &   \\
		&  &  &  &  &  &  &  & \frac{1}{2} & 1 & -\frac{1}{2} & -1 &  & 1 & \frac{1}{2} & -1 &    \\
		&  &  &  &  &  &  &  & -1 & -\frac{1}{2} & 1 &  & -1 & \frac{1}{2} & 1 & -\frac{1}{2} &   \\
		&  &  &  &  &  &  &  & 1 &  & -1 & 1 & -\frac{1}{2} & -\frac{1}{2} & 1 & -\frac{1}{2} &   \\
		&  &  &  &  &  &  &  & -1 & \frac{1}{2} &  & -\frac{1}{2} & 1 & -1 & 1 & -\frac{1}{2} &    \\
		&  &  &  &  &  &  &  & 1 & -1 & 1 & -1 & \frac{1}{2} & -\frac{1}{2} & \frac{1}{2} &  &    \\
		&  &  &  &  &  &  &  &  & -\frac{1}{2} & -\frac{1}{2} & -\frac{1}{2} & -1 & -1 & -1 & -1 &   \\
	\end{rsmallmatrix}\hspace{-0.1cm}\right],
\end{equation*}

\begin{equation*}
	\footnotesize
	{\mathbf{L}}_2=
	\left[\hspace{-0.1cm}\begin{rsmallmatrix}\\
		-\frac{1}{2} & -1 & -1 & -1 & -1 & -\frac{1}{2} &  & \frac{1}{2} & 1 & 1 & 1 & \frac{1}{2} &  & -\frac{1}{2} & -1 & -1 \\
		\frac{1}{2} & 1 & 1 & 1 &  & -\frac{1}{2} & -1 & -1 & -\frac{1}{2} & \frac{1}{2} & 1 & 1 & 1 &  & -\frac{1}{2} & -1 \\
		-\frac{1}{2} & -1 & -1 &  & 1 & 1 & \frac{1}{2} & -\frac{1}{2} & -1 & -1 &  & 1 & 1 & \frac{1}{2} & -\frac{1}{2} & -1 \\
		\frac{1}{2} & 1 & \frac{1}{2} & -\frac{1}{2} & -1 & -\frac{1}{2} & 1 & 1 & \frac{1}{2} & -1 & -1 &  & 1 & 1 &  & -1 \\
		-\frac{1}{2} & -1 &  & 1 & \frac{1}{2} & -1 & -1 &  & 1 & \frac{1}{2} & -1 & -1 & \frac{1}{2} & 1 & \frac{1}{2} & -1 \\
		1 & 1 & -\frac{1}{2} & -1 & \frac{1}{2} & 1 &  & -1 &  & 1 & \frac{1}{2} & -1 & -\frac{1}{2} & 1 & \frac{1}{2} & -1 \\
		-1 & -\frac{1}{2} & 1 & \frac{1}{2} & -1 & -\frac{1}{2} & 1 &  & -1 &  & 1 & -\frac{1}{2} & -1 & \frac{1}{2} & 1 & -1 \\
		1 & \frac{1}{2} & -1 & \frac{1}{2} & 1 & -1 & -\frac{1}{2} & 1 &  & -1 & 1 & \frac{1}{2} & -1 &  & 1 & -\frac{1}{2} \\
		-1 &  & 1 & -1 &  & 1 & -1 & -\frac{1}{2} & 1 & -1 & -\frac{1}{2} & 1 & -\frac{1}{2} & -\frac{1}{2} & 1 & -\frac{1}{2} \\
		1 & -\frac{1}{2} & -\frac{1}{2} & 1 & -1 &  & 1 & -1 & \frac{1}{2} & \frac{1}{2} & -1 & 1 &  & -1 & 1 & -\frac{1}{2} \\
		-1 & \frac{1}{2} &  & -1 & 1 & -1 & \frac{1}{2} & \frac{1}{2} & -1 & 1 & -\frac{1}{2} &  & 1 & -1 & 1 & -\frac{1}{2} \\
		1 & -1 & \frac{1}{2} &  & -\frac{1}{2} & 1 & -1 & 1 & -\frac{1}{2} &  & \frac{1}{2} & -1 & 1 & -1 & 1 & -\frac{1}{2} \\
		-1 & 1 & -1 & \frac{1}{2} & -\frac{1}{2} &  & \frac{1}{2} & -\frac{1}{2} & 1 & -1 & 1 & -1 & 1 & -1 & \frac{1}{2} &  \\
		1 & -1 & 1 & -1 & 1 & -1 & 1 & -1 & 1 & -\frac{1}{2} & \frac{1}{2} & -\frac{1}{2} & \frac{1}{2} & -\frac{1}{2} &  &  \\
		&  & -\frac{1}{2} & -\frac{1}{2} & -\frac{1}{2} & -\frac{1}{2} & -\frac{1}{2} & -1 & -1 & -1 & -1 & -1 & -1 & -1 & -1 & -1 \\
		& \frac{1}{2} & 1 & 1 & 1 & 1 & 1 & 1 & \frac{1}{2} & \frac{1}{2} &  & -\frac{1}{2} & -\frac{1}{2} & -1 & -1 & -1 \\
	\end{rsmallmatrix}\hspace{-0.1cm}\right].
\end{equation*}

\subsection{Matrix Factorization for $N=64$}
The low-complexity matrix $\mathbf{T}_{64,1}$ can be represented as:
\begin{eqnarray*}
	\mathbf{T}_{64,1}=\mathbf{P}_{64}
	\cdot
	\left[
	\begin{array}{ccc}
		\mathbf{Z}_{1}&&\\
		&{\mathbf{Z}}_{2}&\\
		&&{\mathbf{Z}}_{3}
	\end{array}
	\right]
	\cdot
	\left[
	\begin{array}{cc}
		\mathbf{B}_{2}&\\
		&{\mathbf{I}}_{62}
	\end{array}
	\right]
	\cdot
	\left[
	\begin{array}{cc}
		\mathbf{B}_{4}&\\
		&{\mathbf{I}}_{60}
	\end{array}
	\right]
	\cdot
	\left[
	\begin{array}{cc}
		\mathbf{B}_{8}&\\
		&{\mathbf{I}}_{56}
	\end{array}
	\right] \\
	\cdot
	\left[
	\begin{array}{cc}
		\mathbf{B}_{16}&\\
		&{\mathbf{I}}_{48}
	\end{array}
	\right]
	\cdot
	\left[
	\begin{array}{cc}
		\mathbf{B}_{32}&\\
		&{\mathbf{I}}_{32}
	\end{array}
	\right]
	\cdot
	\mathbf{B}_{64}
\end{eqnarray*}
where
\begin{equation*}
	\begin{split}	{\mathbf{P}}_{64}=&\scalebox{0.9}{(1\;32\;17\;22\;42\;37\;27\;62\;5\;40\;33\;19\;30\;14\;4\;24\;50\;53\;59\;9\;28\;2\;16\;18\;26\;58\;7\;8\;20\;34\;21\;38\;29\;6\;56)} \\
		&\scalebox{0.9}{(3\;48\;49\;51\;55\;63\;13\;60\;11\;44\;41\;35\;23\;46\;45\;43\;39\;31\;10\;36\;25\;54\;61\;15\;12\;52\;57)},
	\end{split}
\end{equation*}

\begin{equation*}
	\footnotesize
	{\mathbf{Z}}_1=
	\left[\hspace{-0.1cm}\begin{rsmallmatrix}\\
		1 &  &  &  &  &  &  &  &  &  &  &  &  &  &  & \\
		&-1& &  &  &  &  &  &  &  &  &  &  &  &  &  & \\
		&  &-1&-1& &  &  &  &  &  &  &  &  &  &  &  & \\
		&  & 1&  & &  &  &  &  &  &  &  &  &  &  &  & \\
		&  &  &  & 1&1&  & -1& &  &  &  &  &  &  &  & \\
		&  &  &  &-1& & 1& -1& &  &  &  &  &  &  &  & \\
		&  &  &  & 1&-1&1&   & &  &  &  &  &  &  &  & \\
		&  &  &  &  &-1&-1&-1& &  &  &  &  &  &  &  & \\
		&  &  &  &  &  &  &  & -1 & -1 & -1 &  & 1 & 1 &  & -1\\
		&  &  &  &  &  &  &  & 1 & 1 &  & -1 &  & 1 & 1 & -1 \\
		&  &  &  &  &  &  &  & -1 & -1 & 1 &  & -1 &  & 1 & -1\\
		&  &  &  &  &  &  &  & 1 &  & -1 & 1 &  & -1 & 1 & -1 \\
		&  &  &  &  &  &  &  & -1 & 1 &  & -1 & 1 & -1 & 1 &  \\
		&  &  &  &  &  &  &  & 1 & -1 & 1 & -1 & 1 & -1 &  &  \\
		&  &  &  &  &  &  &  &  &   & -1&-1 & -1& -1& -1& -1 \\
		&  &  &  &  &  &  &  &  & 1 & 1 & 1 & 1 &  & -1 & -1 \\
	\end{rsmallmatrix}\hspace{-0.1cm}\right],\vspace{0.3cm} \\
\end{equation*}

\begin{equation*}
	\footnotesize
	{\mathbf{Z}}_2=
	\left[\hspace{-0.1cm}\begin{rsmallmatrix}
		-1 & -1 & -1 &  & 1 & 1 & 1 &  & -1 & -1 &  & 1 & 1 & 1 &  & -1\\
		1 & 1 & 1 & -1 & -1 &  & 1 & 1 &  & -1 & -1 &  & 1 & 1 &  & -1\\
		-1 & -1 &  & 1 & 1 & -1 & -1 &  & 1 & 1 & -1 & -1 &  & 1 &  & -1\\
		1 & 1 & -1 & -1 &  & 1 &  & -1 &  & 1 &  & -1 & -1 & 1 & 1 & -1\\
		-1 & -1 & 1 & 1 & -1 &  & 1 &  & -1 &  & 1 &  & -1 & 1 & 1 & -1 \\
		1 &  & -1 &  & 1 & -1 & -1 & 1 &  & -1 & 1 & 1 & -1 &  & 1 & -1\\
		-1 &  & 1 & -1 &  & 1 & -1 &  & 1 & -1 &  & 1 & -1 & -1 & 1 & -1\\
		1 &  & -1 & 1 & -1 &  & 1 & -1 &  & 1 & -1 & 1 &  & -1 & 1 & -1\\
		-1 & 1 &  & -1 & 1 & -1 &  & 1 & -1 & 1 & -1 &  & 1 & -1 & 1 & \\
		1 & -1 &  &  & -1 & 1 & -1 & 1 & -1 &  & 1 & -1 & 1 & -1 & 1 & \\
		-1 & 1 & -1 & 1 &  &  &  & -1 & 1 & -1 & 1 & -1 & 1 & -1 & 1 & \\
		1 & -1 & 1 & -1 & 1 & -1 & 1 & -1 & 1 & -1 & 1 & -1 &  &  &  & \\
		&  &  &  & -1 & -1 & -1 & -1 & -1 & -1 & -1 & -1 & -1 & -1 & -1 & -1\\
		& 1 & 1 & 1 & 1 & 1 & 1 & 1 & 1 &  &  &  & -1 & -1 & -1 & -1\\
		& 1 & 1 & 1 &  & -1 & -1 & -1 & -1 &  & 1 & 1 & 1 &  & -1 & -1\\
		& -1 & -1 & -1 & -1 & -1 &  & 1 & 1 & 1 & 1 & 1 &  &  & -1 & -1\\
	\end{rsmallmatrix}\hspace{-0.1cm}\right],\vspace{0.3cm}\\
	{\mathbf{Z}}_{3}=
	\left[
	\begin{array}{ccc}
		\mathbf{Q}_{1}&\mathbf{Q}_{2}\\
		\mathbf{Q}_{3}&{\mathbf{Q}}_{4}
	\end{array}
	\right],
\end{equation*}

and
\begin{equation*}
	\footnotesize
	{\mathbf{Q}}_{1}=
	\left[\hspace{-0.1cm}\begin{rsmallmatrix}
		-1 & -1 & -1 &   & 1 & 1 & 1 &   & -1 & -1 & -1 &   & 1 & 1 & 1 &   \\
		1 & 1 & 1 &   & -1 & -1 &   & 1 & 1 & 1 &   & -1 & -1 &   & 1 & 1 \\
		-1 & -1 & -1 & 1 & 1 & 1 & -1 & -1 & -1 &   & 1 & 1 &   & -1 & -1 &   \\
		1 & 1 &   & -1 & -1 &   & 1 & 1 &   & -1 & -1 & 1 & 1 & 1 & -1 & -1 \\
		-1 & -1 &   & 1 & 1 & -1 & -1 &   & 1 & 1 &   & -1 &   & 1 & 1 &   \\
		1 & 1 &   & -1 &   & 1 & 1 & -1 & -1 &   & 1 & 1 & -1 & -1 &   & 1 \\
		-1 & -1 & 1 & 1 &   & -1 &   & 1 & 1 & -1 & -1 & 1 & 1 &   & -1 &   \\
		1 & 1 & -1 & -1 & 1 & 1 & -1 & -1 &   & 1 &   & -1 &   & 1 &   & -1 \\
		-1 & -1 & 1 & 1 & -1 & -1 & 1 & 1 & -1 &   & 1 &   & -1 &   & 1 &   \\
		1 & 1 & -1 &   & 1 &   & -1 &   & 1 & -1 & -1 & 1 & 1 & -1 &   & 1 \\
		-1 & -1 & 1 &   & -1 & 1 & 1 & -1 &   & 1 &   & -1 & 1 & 1 & -1 &   \\
		1 &   & -1 & 1 & 1 & -1 &   & 1 & -1 & -1 & 1 &   & -1 & 1 & 1 & -1 \\
		-1 &   & 1 & -1 &   & 1 & -1 & -1 & 1 &   & -1 & 1 &   & -1 & 1 &   \\
		1 &   & -1 & 1 &   & -1 & 1 &   & -1 & 1 &   & -1 & 1 &   & -1 & 1 \\
		-1 &   & 1 & -1 & 1 & 1 & -1 & 1 &   & -1 & 1 &   & -1 & 1 & -1 &   \\
		1 &   & -1 & 1 & -1 &   & 1 & -1 & 1 &   & -1 & 1 & -1 &   & 1 & -1 \\
	\end{rsmallmatrix}\hspace{-0.1cm}\right],
\end{equation*}
\begin{equation*}
	\footnotesize
	{\mathbf{Q}}_{2}=
	\left[\hspace{-0.1cm}\begin{rsmallmatrix}
		-1 & -1 &   & 1 & 1 & 1 &   & -1 & -1 & -1 &   & 1 & 1 & 1 &   & -1 \\
		& -1 & -1 & -1 &   & 1 & 1 &   & -1 & -1 & -1 & 1 & 1 & 1 &   & -1 \\
		1 & 1 &   & -1 & -1 &   & 1 & 1 &   & -1 & -1 &   & 1 & 1 &   & -1 \\
		& 1 & 1 &   & -1 & -1 &   & 1 & 1 & -1 & -1 &   & 1 & 1 &   & -1 \\
		-1 & -1 & 1 & 1 &   & -1 & -1 & 1 & 1 &   & -1 & -1 & 1 & 1 &   & -1 \\
		& -1 & -1 & 1 & 1 &   & -1 &   & 1 & 1 & -1 & -1 &   & 1 & 1 & -1 \\
		1 &   & -1 & -1 & 1 & 1 & -1 & -1 &   & 1 &   & -1 &   & 1 & 1 & -1 \\
		& 1 &   & -1 &   & 1 &   & -1 & -1 & 1 & 1 & -1 & -1 & 1 & 1 & -1 \\
		-1 &   & 1 &   & -1 &   & 1 &   & -1 & 1 & 1 & -1 & -1 & 1 & 1 & -1 \\
		& -1 &   & 1 & -1 & -1 & 1 & 1 & -1 &   & 1 &   & -1 & 1 & 1 & -1 \\
		1 &   & -1 & 1 & 1 & -1 &   & 1 & -1 & -1 & 1 &   & -1 &   & 1 & -1 \\
		& 1 & -1 &   & 1 &   & -1 & 1 &   & -1 & 1 & 1 & -1 &   & 1 & -1 \\
		-1 & 1 & 1 & -1 & 1 & 1 & -1 &   & 1 & -1 &   & 1 & -1 &   & 1 & -1 \\
		& -1 & 1 &   & -1 & 1 &   & -1 & 1 & -1 & -1 & 1 & -1 & -1 & 1 & -1 \\
		1 & -1 &   & 1 & -1 &   & 1 & -1 & 1 &   & -1 & 1 &   & -1 & 1 & -1 \\
		& 1 & -1 & 1 &   & -1 & 1 & -1 &   & 1 & -1 & 1 &   & -1 & 1 & -1 \\
	\end{rsmallmatrix}\hspace{-0.1cm}\right],
\end{equation*}

\begin{equation*}
	\footnotesize
	{\mathbf{Q}}_{3}=
	\left[\hspace{-0.1cm}\begin{rsmallmatrix}\\
		-1 & 1 &   & -1 & 1 & -1 &   & 1 & -1 & 1 &   & -1 & 1 & -1 &   & 1 \\
		1 & -1 &   & 1 & -1 & 1 & -1 &   & 1 & -1 & 1 & -1 &   & 1 & -1 & 1 \\
		-1 & 1 &   &   & 1 & -1 & 1 & -1 &   & 1 & -1 & 1 & -1 & 1 &   & -1 \\
		1 & -1 & 1 &   & -1 & 1 & -1 & 1 & -1 & 1 &   &   & 1 & -1 & 1 & -1 \\
		-1 & 1 & -1 &   &   &   & 1 & -1 & 1 & -1 & 1 & -1 & 1 &   &   & 1 \\
		1 & -1 & 1 & -1 & 1 &   &   &   & -1 & 1 & -1 & 1 & -1 & 1 & -1 & 1 \\
		-1 & 1 & -1 & 1 & -1 & 1 & -1 & 1 &   &   &   &   &   & -1 & 1 & -1 \\
		1 & -1 & 1 & -1 & 1 & -1 & 1 & -1 & 1 & -1 & 1 & -1 & 1 & -1 & 1 & -1 \\
		&   &   &   &   &   &   &   & -1 & -1 & -1 & -1 & -1 & -1 & -1 & -1 \\
		&   &   & 1 & 1 & 1 & 1 & 1 & 1 & 1 & 1 & 1 & 1 & 1 & 1 & 1 \\
		& 1 & 1 & 1 & 1 & 1 & 1 & 1 &   &   & -1 & -1 & -1 & -1 & -1 & -1 \\
		& -1 & -1 & -1 & -1 & -1 &   &   & 1 & 1 & 1 & 1 & 1 &   &   & -1 \\
		& 1 & 1 & 1 & 1 &   &   & -1 & -1 & -1 & -1 &   & 1 & 1 & 1 & 1 \\
		& 1 & 1 & 1 &   & -1 & -1 & -1 &   & 1 & 1 & 1 &   & -1 & -1 & -1 \\
		&   & -1 & -1 & -1 & -1 & -1 & -1 & -1 & -1 & -1 &   &   &   & 1 & 1 \\
		& -1 & -1 & -1 &   &   & 1 & 1 & 1 &   & -1 & -1 & -1 & -1 &   & 1 \\
	\end{rsmallmatrix}\hspace{-0.1cm}\right],
\end{equation*}

\begin{equation*}
	\footnotesize
	{\mathbf{Q}}_{4}=
	\left[\hspace{-0.1cm}\begin{rsmallmatrix}\\
		-1 & 1 & -1 &   & 1 & -1 & 1 &   & -1 & 1 & -1 &   & 1 & -1 & 1 &   \\
		-1 &   & 1 & -1 & 1 & -1 &   & 1 & -1 & 1 &   &   & 1 & -1 & 1 &   \\
		1 & -1 & 1 & -1 &   & 1 & -1 & 1 & -1 &   &   & -1 & 1 & -1 & 1 &   \\
		1 &   &   & 1 & -1 & 1 & -1 & 1 &   &   & 1 & -1 & 1 & -1 & 1 &   \\
		-1 & 1 & -1 & 1 & -1 & 1 &   &   & 1 & -1 & 1 & -1 & 1 & -1 & 1 &   \\
		-1 & 1 &   &   &   & -1 & 1 & -1 & 1 & -1 & 1 & -1 & 1 & -1 &   &   \\
		1 & -1 & 1 & -1 & 1 & -1 & 1 & -1 & 1 & -1 & 1 & -1 & 1 &   &   &   \\
		1 & -1 & 1 & -1 & 1 & -1 & 1 & -1 &   &   &   &   &   &   &   &   \\
		-1 & -1 & -1 & -1 & -1 & -1 & -1 & -1 & -1 & -1 & -1 & -1 & -1 & -1 & -1 & -1 \\
		1 & 1 & 1 &   &   &   &   &   & -1 & -1 & -1 & -1 & -1 & -1 & -1 & -1 \\
		-1 &   &   & 1 & 1 & 1 & 1 & 1 & 1 & 1 &   &   &   & -1 & -1 & -1 \\
		-1 & -1 & -1 & -1 &   &   & 1 & 1 & 1 & 1 & 1 & 1 &   & -1 & -1 & -1 \\
		1 &   & -1 & -1 & -1 & -1 & -1 &   & 1 & 1 & 1 & 1 &   &   & -1 & -1 \\
		-1 &   & 1 & 1 & 1 &   & -1 & -1 & -1 &   & 1 & 1 & 1 &   & -1 & -1 \\
		1 & 1 & 1 & 1 & 1 & 1 & 1 & 1 &   &   &   & -1 & -1 & -1 & -1 & -1 \\
		1 & 1 & 1 &   & -1 & -1 & -1 & -1 &   & 1 & 1 & 1 & 1 &   & -1 & -1 \\
	\end{rsmallmatrix}\hspace{-0.1cm}\right].
\end{equation*}

\onecolumn

{\small
\singlespacing

\bibliographystyle{siam}

\bibliography{references}
}

\end{document}